\documentclass[11pt]{article}
\pdfoutput=1 

\usepackage[margin=1.4in]{geometry}
\usepackage[authoryear]{natbib}

\setcitestyle{authoryear,open={(},close={)}}

\input epsf.sty

\usepackage{amsmath,amsfonts,amssymb,amsthm}
\usepackage{float,dsfont,url,color}
\usepackage{graphicx}
\usepackage{bbm}
\usepackage{color}
\usepackage{amsmath}
\usepackage{authblk}
\usepackage{siunitx}

\title{Bayesian detection of piecewise linear trends in replicated time-series with application to growth data modelling}
\author[1]{Panagiotis Papastamoulis\thanks{\tt papastamoulis@aueb.gr}}
\author[2]{Takanori Furukawa}
\author[2]{Norman Van Rhijn}
\author[2]{Michael Bromley}
\author[2]{Elaine Bignell}
\author[3]{Magnus Rattray}
\affil[1]{Department of Statistics, Athens University of Economics and Business, Greece}
\affil[2]{Division of Infection, Immunity \& Respiratory Medicine, University of Manchester, UK}
\affil[3]{Division of Informatics, Imaging \& Data Sciences, Faculty of Biology, Medicine and Health, University of Manchester, UK}

\date{}
\begin{document}
\maketitle

\begin{abstract}
We consider the situation where a temporal process is composed of contiguous segments with differing slopes and replicated noise-corrupted time series measurements are observed. The unknown mean of the data generating process is modelled as a piecewise linear function of time with an unknown number of change-points. We develop a Bayesian approach to infer the joint posterior distribution of the number and position of change-points as well as the unknown mean parameters. A-priori, the proposed model uses an overfitting number of mean parameters but, conditionally on a set of change-points, only a subset of them influences the likelihood. An exponentially decreasing prior distribution on the number of change-points gives rise to a posterior distribution concentrating on sparse representations of the underlying sequence. A Metropolis-Hastings Markov chain Monte Carlo (MCMC) sampler is constructed for approximating the posterior distribution. Our method is benchmarked using simulated data and is applied to uncover differences in the dynamics of fungal growth from imaging time course data collected from different strains. The source code is available on CRAN.
\end{abstract}

\section{Introduction}
\label{s:intro}

In many applications a non-stationary time series consists of an unknown number of segments. The observed data is described by different statistical generative models within each segment. In such cases, the objective is to identify the number and position of change-points which give rise to different segments of the data as well as to infer all remaining parameters of the underlying statistical model. In principle, there are two approaches for answering these questions: online and offline segmentation \citep{basseville1993detection}, which refer to the task of inferring changes during or after the observation process, respectively. In this work, the latter scenario is considered. 

We develop a Bayesian method for detecting an unknown number of change-points in the slope of multiple replicated time series. There are many methods for detecting change-points, with the majority of them focused on analysis of univariate time series (reviewed below). Our problem differs from these as we model changes in slope rather than fitting a step function to the mean, and we consider multiple time series with replication. For a given period $t = 1,\ldots,T$ we observe multiple time series which are assumed independent, each one consisting of multiple measurements (replicates). Each time series is assumed to have its own segmentation, which is common among its replicates. Thus, different time series have distinct mean parameters in the underlying normal distribution.

\begin{figure}
\centering
\includegraphics[scale=0.5]{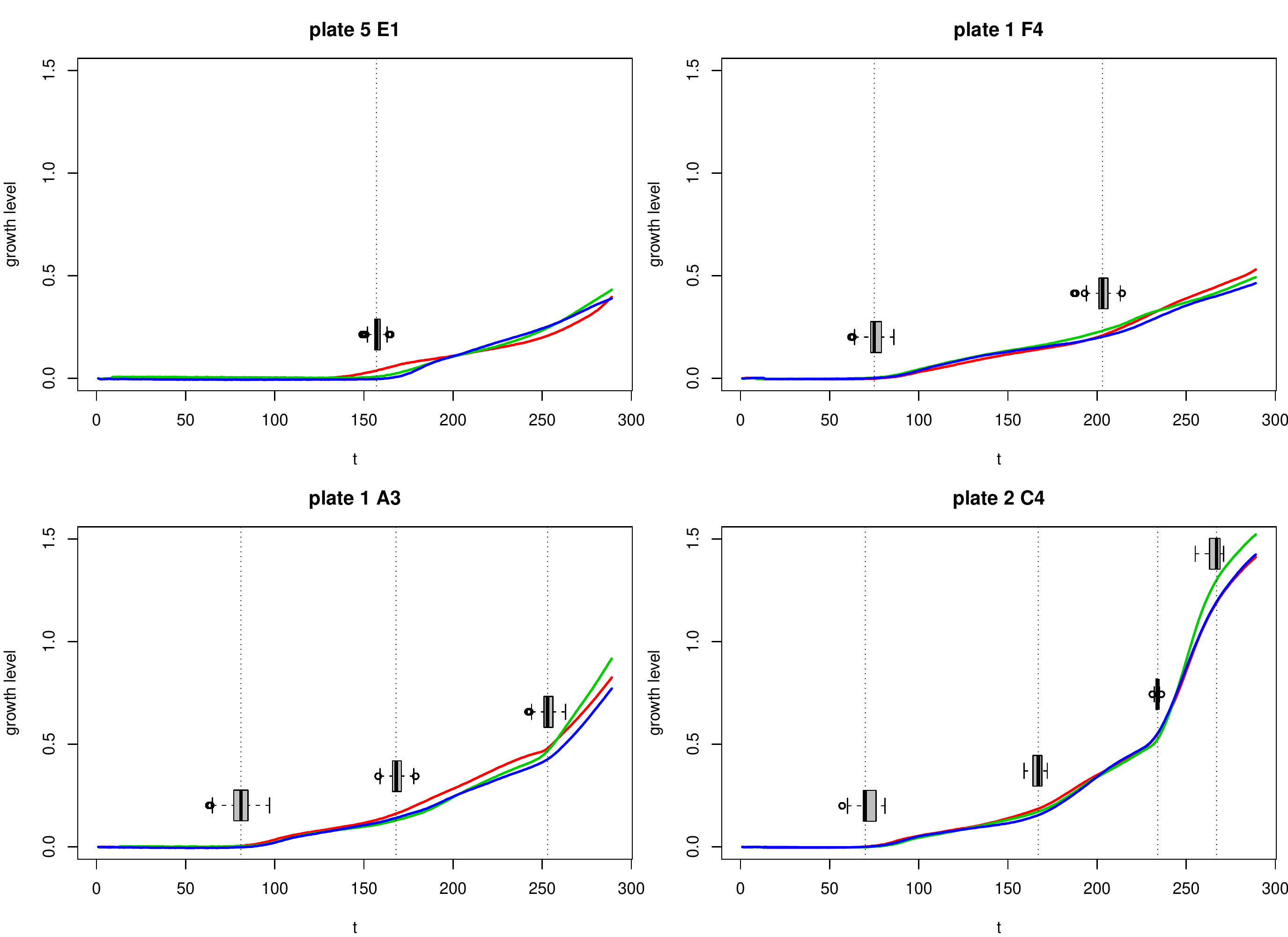}
\caption{A subset of four growth data series described in Section \ref{sec:growth}, consisting of the growth level of three replicates (red, green and blue), measured every 10 minutes for a series of $T = 289$ time-points. The boxplots display the marginal posterior distribution of each change-point, conditionally on the inferred Maximum A Posteriori number of change-points according the MCMC sampler detailed in Section \ref{sec:mcmc}.}
\label{fig:timeCourse}
\end{figure}

Our method is motivated by the need to analyse fungal growth attributes on a massively parallel scale. Specifically, we are interested in identifying mutations which affect fitness in the major human fungal pathogen \textit{Aspergillus fumigatus}. In such studies, growth is characterized by different phases which can be reasonably described by a piecewise linear model, as illustrated in Figure \ref{fig:timeCourse}. However, the protocol is more widely relevant to any scenario in which a detailed characterisation of microbial growth attributes is required. Microbial growth is a complex characteristic which is heavily influenced by nutritional, metabolic, proliferative, physiological and genetic factors. Multiple techniques have been developed with which to quantify microbial growth, including direct quantitation of cell counts using flow cytometry or microscopy, colony counts, biomass quantitation, or indirect methods involving light scattering or turbidity measurement in liquid phase cultures, or dye-based methods. Optimisation of data acquisition and analysis has received rather less attention, particularly where the quantitation of growth characteristics in filamentous and aggregative microorganisms, such as \textit{A.~fumigatus} or \textit{Streptomyces coelicolor} is complicated by the occurrence of one or several morphological shifts during the mitotic life cycle \citep{fischer2013universally} leading to altered light scattering patterns dependent upon the size and shape of the particulate sample (bacteria or yeast), as well as difference in the index of refraction between the particles and the culture media \citep{stevenson2016general}. In the latter instance an accurate means of defining the number and timing of change-points during growth curve analysis would  significantly empower the optimisation of drug discovery screens where inhibitors of microbial growth might be sought; or in optimisation of biotechnological processes where moderation of microbial growth conditions to favour a particular growth phase might boost industrial production of enzymes or metabolites.

In the seminal paper of \cite{green1995reversible}, the Reversible Jump MCMC (RJMCMC) algorithm used to detect the number of change-points in  coal mining disaster data. Subsequently, the RJMCMC methodology was  applied to a variety of change-point detection problems \citep{984776, doi:10.1093/biostatistics/4.1.143, BIOM:BIOM1328, zhao2010bayesian}. \cite{lavielle2001application} proposed an MCMC sampler to estimate the number of change-points by introducing a latent sequence of independent and identically distributed Bernoulli random variables $r_t$; $t = 1,\ldots,T$, with $T$ denoting the number of time-points. In this context, $r_j = 1$  indicates that a change occurs at  time $t = j$, while $r_j = 0$  means that no change occurs. This  approach has the advantage that it can infer the target posterior distribution using an MCMC sampler that operates on random variables of constant dimension, in contrast to the RJMCMC approach. However, it turns out that the estimated marginal posterior  probabilities of these artificial binary random variables overestimates the true number of change-points. We illustrate that similar issues arise in our set-up when imposing typical prior assumptions on the number of change-points, such as a Poisson distribution. To overcome this problem, \cite{lavielle2001application} inferred configurations of change-points of high probability by sampling from a modified posterior distribution which is a tempered version of the original target, using a simulated annealing MCMC algorithm. In our set-up, we demonstrate that we are able to accurately infer the number of change-points when using priors that heavily penalize large values of change-points \citep{castillo2012}. 

\cite{Chib1998221} formulates the change-point model in terms of a latent discrete state variable corresponding to the regime from which a particular observation has been drawn. The posterior distribution for a given number of change-points is then approximated using MCMC sampling, while inference on the number of change-points is carried out by estimating the marginal likelihood of the model using the method in \cite{chib1995marginal}. \cite{fearnhead2006exact} discusses exact Bayesian inference by assuming that the joint posterior distribution of the parameters is independent across the segments of the 
time series and also presents an extension that allows the signal to perform a random walk within each segment.  Other Bayesian methods include \citet{dobigeon2007joint, hutter2007,Kim2010,RSSC:RSSC715,
PhysRevE.84.021120,
doi:10.1080/03610926.2016.1161797, Schwaller2017}. 
In all of the aforementioned studies, change-points are defined via a step-function in the mean, assuming that segment parameters are independent. This is not the case in our approach, where we are explicitly imposing a continuity assumption between segments which allows us to detect changes in slope.  With the exception of \citet{dobigeon2007joint, Schwaller2017}, all other methods are focusing on univariate time-series.

There are relatively few studies looking specifically at a change-in-slope model \citep{Schroeder, 1748-9326-10-8-084002, baranowski2016narrowest, maidstone2017detecting}, although they only consider a single time-series. Popular non-Bayesian methods such as binary segmentation \citep{10.2307/2529204, fryzlewicz2014wild} do not work for detecting changes-in-slope \citep{baranowski2016narrowest}. Furthermore,  standard dynamic programming approaches \citep{jackson2005algorithm, killick2012optimal} cannot be directly applied to our problem as discussed by \citet{maidstone2017detecting}. 


There is a wide range of non-Bayesian approaches to change-point estimation, see for example \citet{BIOM:BIOM903,lu2010,Picard20111160,
doi:10.1080/10618600.2012.674653,
Chamroukhi2013633,RSSB:RSSB12047}. However, we choose a Bayesian approach for its competency in quantifying uncertainty and flexibility for incorporating prior information. 

We construct a Metropolis-Hastings MCMC sampler \citep{metropolis1953equation, 10.2307/2334940} for jointly inferring the number and position of change-points as well as the related mean parameters by adopting ideas from inference over sparse representations of sequences \citep{castillo2012}.  An advantage of our approach is that the proposed MCMC algorithm is straightforward to implement since it is based on standard Metropolis--Hastings move types and demands  small modelling effort compared to other methods. Although exact  integration is possible given the number and locations of change-points, it is time consuming hence we exploit the convenience of the MCMC sampler to approximately sample from the joint posterior distribution of the parameters. The dimensionality of the parameter space is fixed, thus our method avoids the complex step of designing trans--dimensional MCMC transitions as required by RJMCMC methods. Furthermore, we do not have to consider modified versions of the target posterior distribution \citep{lavielle2001application}, 
 and there is no requirement for  fitting the same model under different number of change-points and approximating the marginal likelihood for model selection.

The rest of the paper is organised as follows. Section \ref{sec:model}
 introduces the proposed model and the corresponding prior assumptions are presented in Section \ref{sec:prior}. The main MCMC sampler we use is detailed in Section \ref{sec:mcmc}. Sections \ref{sec:variance} and \ref{sec:unknownVar} discuss variance estimation procedures, depending whether the variance is treated as a known parameter or not. The proposed method is illustrated in simulated and real data in Sections \ref{sec:sim} and \ref{sec:growth}, respectively. The paper concludes in Section \ref{sec:discussion}. More details on the simulation procedure and additional results are given in the Appendix. 

\section{Model}\label{sec:model}
Let $X_{ntr}$ denote a random variable describing replicate $r$ at time point $t$ for time series $n$, $n = 1, \ldots,N$; $t = 1, \ldots,T$; $r = 1,\ldots,R$. It is assumed that $\{X_{ntr};r=1,\ldots,R\}$ is a normally distributed random sample and furthermore that  measurements are independent across time, that is:
\begin{equation}\label{eq:model0}
X_{ntr}\sim
\mathcal N\left(\theta_{nt}, \sigma^2_{nt}\right), \quad \theta_{nt}\in\mathbb R, \sigma^2_{nt}>0
\end{equation}
independent for $n = 1, \ldots,N$; $t = 1, \ldots,T$; $r = 1,\ldots,R$, where $\mathcal N(\cdot,\cdot)$ denotes the normal distribution. At first, we will consider that the variances $\{\sigma^2_{nt}, n = 1,\ldots,N;t=1,\ldots,n\}$ are known. In practice, the variance per time-point is estimated at a pre-processing stage as exemplified in Section \ref{sec:variance}. The general case where the variance is treated as a random variable is addressed in Section \ref{sec:unknownVar}. Without any further assumptions, the parameterization of the normal distributions in Equation \eqref{eq:model0} introduces a large number of mean parameters: a distinct mean parameter $\theta_{nt}\in\Theta = \mathbb{R}$ is assigned to each sample ($n$) and time-point $(t)$. However, $\theta_{nt}$ is shared across replicates ($r$).

For sample $n$ and an unknown non-negative integer $\ell_n\geqslant 0$, assume that there are $\ell_n+1$ underlying phases of mean behaviour, identified by the ordered time points \begin{equation}
\label{eq:constraint}
1< \tau_{n1}  <\tau_{n2}  < \ldots<\tau_{n\ell_n} < T.
\end{equation}  Note that both the elements as well as the length of the ordered $\ell_n$-tuple $\boldsymbol \tau_n=\left(\tau_{n1},\ldots,\tau_{n\ell_n}\right)$ depends on $n$. Assume that for each phase the mean function is linear in time. For phase $j = 1,\ldots,\ell_n+1$, the piecewise linear mean measurement levels are defined as follows:
\begin{equation}\label{eq:mean}
\mu\left(t;\boldsymbol{\theta}_n,\boldsymbol{\tau}_n\right)  =  \theta_{n\tau_{n;j-1}} + \frac{\theta_{n\tau_{nj}}-\theta_{n\tau_{n;j-1}}}{\tau_{nj} - \tau_{n;j-1}}\left(t-\tau_{n;j-1}\right),\quad \tau_{n;j-1} \leqslant t \leqslant \tau_{nj},
\end{equation}
where we also define $\tau_{n0} := 1$ and $\tau_{n;\ell_n+1}:=T$; $\forall n =1,\ldots,N$. Note that Equation \eqref{eq:mean} imposes continuity in the segment means, which is a crucial difference with other approaches (mentioned in the Introduction): the mean for the end of one segment should be equal to the mean at the start of the next segment.

Thus, for sample $n$, conditionally on $\boldsymbol \tau_{n}$, we can write that
\begin{equation}\label{eq:model}
X_{ntr}|\boldsymbol \tau_n\sim
\mathcal N\left(\mu\left(t;\boldsymbol{\theta}_n,\boldsymbol{\tau}_n\right), \sigma^2_{nt}\right),
\end{equation}
independent for $n=1,\ldots,N$; $t = 1,\ldots,T$; $r = 1,\ldots,R$. Note that given $\boldsymbol \tau_{n}$, the likelihood depends on $\boldsymbol \theta$ only through the subset
\begin{eqnarray}\label{eq:thetatau}
\boldsymbol\theta_{\boldsymbol \tau_{n}}&:=& \left\{\theta_{n1}, \theta_{n\tau_{n1}}, \ldots, \theta_{n\tau_{n\ell_n}}, \theta_{nT}\right\}.
\end{eqnarray}
To be precise, the likelihood is defined as 
\begin{equation}
f\left(\boldsymbol x_n|\boldsymbol\theta_{\boldsymbol\tau_n}, \boldsymbol\sigma^2_n\right) =\left[\prod_{r=1}^{R}\varphi
\left(
x_{n1r};\theta_{n1},\sigma^2_{n1}
\right)\right] \prod_{j=0}^{\ell_n}\prod_{t = \tau_{nj}+1}^{\tau_{n;j+1}}\prod_{r=1}^{R}\varphi\left(x_{ntr};\mu(t;\boldsymbol\theta_n,\boldsymbol\tau_n),\sigma^{2}_{nt}\right),
\end{equation}
where $\varphi(\cdot;\mu,\sigma^2)$ denotes the probability density function of the normal distribution with mean $\mu$ and variance $\sigma^2$ and $\mu(t;\boldsymbol\theta_n,\boldsymbol\tau_n)$ defined in Equation \eqref{eq:mean}.

\subsection{Prior assumptions}\label{sec:prior}

For the mean parameters we assume that 
\begin{equation}\label{eq:thetaPrior}
\theta_{nt} \sim \mathcal N\left( \mu_{0t},\frac{\sigma^2_{nt}}{\nu_0}\right)
\end{equation}
independent for $n=1,\ldots,N$; $t=1,\ldots,T$. The quantities $\nu_0 > 0$ and $\mu_{0t}\in\mathbb R$; $t=1,\ldots,T$, correspond to fixed hyper-parameters. The following default values are considered: $\mu_{0t} = \frac{1}{NR}\sum_{n=1}^{N}\sum_{r = 1}^{R}x_{ntr}$, for $t = 1,\ldots,T$. It is suggested that the parameter $\nu_0$ should be sufficiently small so that the prior distribution in \eqref{eq:thetaPrior} has large variability around the global mean. This is particularly important in cases where the multiple time series of an experiment exhibit strong heterogeneity. Values between  $5\times10^{-2}\leqslant \nu_0 \leqslant 5\times 10^{-1}$ performed reasonably well in our setup. The variance $\sigma^2_{nt}$  may be fixed (see Section \ref{sec:variance}) or not (see Section \ref{sec:unknownVar}).

In order to specify the prior distribution of locations for a given number of change-points ($\ell_n$), we are taking into account the prior assumption that later time-points are more likely to contain changes than earlier time-points. The growth levels of $\ell_n$ consecutive time-points during the end of the observation period are more likely to break collinearity than earlier stages. For example, all time-series consist of an initial period with very small and almost constant growth level in which no change is expected to occur. On the other hand, the last part of the observation period may exhibit larger heterogeneity. A simple and efficient way to incorporate such a prior information while supporting (with positive probability) all possible configurations with respect to constraint \eqref{eq:constraint} is the following.

Let $g(i;i_i,i_2) = \frac{1}{i_2-i_1 + 1}\mathbb I(i_1\leqslant i \leqslant i_2)$ denote the probability mass function of the discrete uniform distribution defined over the finite set of integers $i$ such that $i_1 \leqslant i \leqslant i_2$, where $\mathbb I(\cdot)$ denotes the indicator function. For $j=1$ we assume that $\tau_{n1}\sim g(\cdot;2,T-\ell_n)$. For $j\geqslant 2$ and conditionally on the event $(\tau_{n1}=t_{1}, \ldots,\tau_{n;j-1} = t_{j-1})$, we assume that $\tau_{nj}$ follows a (discrete) uniform prior distribution defined over $t_{j-1}+1\leqslant\tau_{nj}\leqslant T-\ell_n+j-1$. Thus, the prior distribution $f\left(\boldsymbol\tau_n|\ell_n>0\right)$ for a specific realization $ (t_1,\ldots,t_{\ell_n})$ of $\boldsymbol\tau_n$ is defined as
\begin{align}\nonumber
f(t_1,\ldots,t_{\ell_n}|\ell_n > 0) = \mathbb{P}\left(\tau_{n1} = t_1,\ldots,\tau_{n\ell_n}=t_{\ell_n}|\ell_n > 0\right)\\
\nonumber
 = \mathbb{P}(\tau_{n1} = t_1)\prod_{j=2}^{\ell_n}\mathbb P(\tau_{nj} = t_j|\tau_{n;j-1} = t_{j-1},\ldots,\tau_{n1} = t_1)\\
\nonumber
= g(t_1;2,T-\ell_n)\prod_{j=2}^{\ell_n}g(t_j;t_{j-1}+1,T-\ell_n+j-1)\\
= \frac{1}{T-\ell_n-1}\prod_{j=2}^{\ell_n}\frac{1}{T-\ell_n+j-t_{j-1}-1}\mathbb I(1< t_1 < \ldots<t_{\ell_n}< T).\label{eq:prior1}
\end{align}

Note that according to Equation \eqref{eq:prior1}, $\mathbb{P}(\tau_{n1} = T-\ell_n,\tau_{n2}=T-\ell_n+1,\ldots,\tau_{n\ell_n} = T-1 | \ell_n)\approx 1/T$, while $\mathbb{P}(\,\tau_{n1}=2,\tau_{n2}=3,\ldots,\tau_{n\ell_n}=\ell_n+1|\ell_n)\approx 1/T^{\ell_n}$, which satisfies our prior expectation discussed in the previous paragraph. We assume a-priori independence of $\boldsymbol{\tau}_n$ for $n = 1,\ldots,N$. Obviously, $\boldsymbol{\tau}_n$ makes sense only in the case that the total number of change-points is strictly positive, thus Equation \eqref{eq:prior1} is defined conditionally on the event $\ell_n > 0$. 


Finally, the prior distribution of the number of change-points should be defined. Recall that the distribution in Equation \eqref{eq:model0} assigns a distinct mean parameter per time-point. However, given $\ell_n$, only a small subset of $\theta$'s will influence the likelihood in Equation \eqref{eq:model} and the rest of them will  affect the posterior solely due to their contribution to the prior distribution. Our motivation is based on the fact that we are trying to find very minimal models with few change points because it makes the data easier to interpret. 

For a given number of change-points ($\ell$), the number of possible models equals to $T^*\choose \ell$, a number which grows rapidly with $\ell$, where $T^*=T-2$. In order to penalize model complexity we consider prior distributions which are biased towards sparse configurations. A natural way to incorporate such information on our prior distribution is to assume that the prior probability of $\ell$ change-points is inversely  proportional to $T^*\choose \ell$, that is, the number of models of size $\ell$. This approach is also justified by the work of \cite{castillo2012}, where they consider that the number of change-points follows an \textit{exponentially decreasing} prior distribution.


The prior $\mathbb P(\cdot)$ has exponential decrease if, for some constants $C> 0$ and $D < 1$, 
\begin{equation}\label{eq:expDecrease}
\mathbb P(\ell_n = \ell)\leqslant D \mathbb P(\ell_n = \ell - 1), \quad \ell > C \ell_{*},
\end{equation} 
where $\ell_*$ denotes the true value of $\ell_n$. In the context of multivariate normal mean models with an underlying sparse true mean vector, it has been shown \citep{castillo2012} that asymptotically, priors satisfying \eqref{eq:expDecrease} lead to posterior distributions that concentrate on the sparse underlying true generative model. Members of the family of the so-called ``complexity priors'', defined as,
\begin{equation}\label{eq:ellPrior}
f(\ell) = \mathbb P(\ell_n = \ell)\propto e^{-\alpha\ell\log(bT^*/\ell)}, \quad a, b > 0; \ell = 0,1,2,\ldots
\end{equation}
have exponential decrease \eqref{eq:expDecrease} for $b > 1+e$ \citep{castillo2012}. In our applications we consider the choices $b = 3.72$  and $\alpha = 2$ as default values, but we also report various prior sensitivity checks in the Appendix.  As noted by \cite{castillo2012} it holds that $e^{\ell\log{(T^*/\ell)}}\leqslant\binom{T^*}{\ell}\leqslant e^{\ell\log{(eT^*/\ell)}}$ implying that \eqref{eq:ellPrior} is inversely proportional to the number of models of size $\ell$. Thus, this choice is suited to the purpose of penalizing model complexity. Note that the right hand side of Equation $\eqref{eq:ellPrior}$ for $\ell=0$ should be perceived as the limit $\lim_{\ell\downarrow 0}e^{-\alpha\ell\log(bT^*/\ell)} = 1$. 

Finally, we mention that typical prior assumptions on the number of change-points (for example a truncated Poisson or a uniform distribution over a pre-specified set of non-negative integer values) tend to overfit the number of change-points. This behaviour is demonstrated in Section \ref{sec:sim} using simulated data with a known number of change-points, as well as in Section \ref{Appsec:pois} of the Appendix using our real dataset.

\section{Inference}\label{sec:inference}

\subsection{Metropolis--Hastings MCMC Sampler}\label{sec:mcmc}

Assume first that the variances $\boldsymbol\sigma=\{\sigma^2_{nt}; n=1,\ldots,N, t=1,\ldots,T\}$ are given. This assumption will be relaxed at the end of this section. Observe that conditionally on the vector $\boldsymbol\sigma^2$,  $\{(\boldsymbol\theta_n,\boldsymbol\ell_n,\boldsymbol\tau_n);n=1,\ldots,N\}$  are a-posteriori independent. Therefore, the inferential procedure breaks down to $N$ independent tasks. The posterior distribution is written as
\begin{eqnarray}\label{eq:fullPosterior}
f\left(\boldsymbol\theta, \boldsymbol\ell, \boldsymbol \tau|\boldsymbol x, \boldsymbol \sigma^2\right) & = & \prod_{n=1}^{N}f\left(\boldsymbol\theta_n, \ell_n, \boldsymbol \tau_n|\boldsymbol x_n, \boldsymbol \sigma^2_n\right)
\\
&\propto& \prod_{n=1}^{N}f\left(\boldsymbol x_n|\boldsymbol\theta_{\boldsymbol \tau_n},\boldsymbol \sigma^2\right)f\left(\boldsymbol\theta_n|\boldsymbol \sigma^2\right)f(\boldsymbol \tau_n|\ell_n)f(\ell_n) \ .
\end{eqnarray}

\noindent Although analytical evaluation of the marginal posterior distribution $f(\boldsymbol\tau, \boldsymbol\ell|\boldsymbol x, \boldsymbol\sigma^2) = \int_\Theta f\left(\boldsymbol\theta, \boldsymbol \tau, \boldsymbol\ell|\boldsymbol x, \boldsymbol \sigma^2\right)\mbox{d}\boldsymbol\theta$ is possible since we use conjugate prior,  the integration cannot be carried out independently within each segment due to the continuity constraint. This fact makes the use of standard dynamic programming approaches not applicable in our set-up and the discrete nature of the sampling problem will make the computation of the involved expressions a time consuming task since the possible combinations of change-points increases rapidly with $T$. Thus, in order to make inference we approximately sample from the  target posterior distribution using a Metropolis-Hastings MCMC sampler that updates $(\boldsymbol\theta,\ell, \boldsymbol\tau)$.

At each step, the state of the chain is updated using four move-types: move 1 updates the number of change-points, move 2 updates the mean parameters by using a random walk proposal centered at the current values, move 3 updates the position of change-points and move 4 updates the subset of mean parameters that are not allocated to a change-point. In each case the proposed move is accepted according to the usual Metropolis-Hastings acceptance ratio, that is, $R_n = \min\{1, \alpha_n\}$ where

\begin{equation}\label{eq:mh}
\alpha_n = \frac{f\left(\boldsymbol\theta'_n, \ell'_n, \boldsymbol \tau'_n|\boldsymbol x_n, \boldsymbol \sigma^2_n\right) \mbox{P}_{\mbox{prop}}\left(\left\{\boldsymbol\theta'_n,\ell'_n,\boldsymbol\tau'_n\right\}\rightarrow \left\{\boldsymbol\theta^{(m)}_n,\ell^{(m)}_n,\boldsymbol\tau^{(m)}_n\right\}\right)}{f\left(\boldsymbol\theta_n, \ell_n, \boldsymbol \tau_n|\boldsymbol x_n, \boldsymbol \sigma^2_n\right) \mbox{P}_{\mbox{prop}}\left(\left\{\boldsymbol\theta_n,\ell_n,\boldsymbol\tau_n\right\}\rightarrow \left\{\boldsymbol\theta'_n,\ell'_n,\boldsymbol\tau'_n\right\} \right)}.
\end{equation}
In Equation \eqref{eq:mh}, $\left(\boldsymbol\theta_n,\ell_n,\boldsymbol\tau_n\right)$ denotes the current state of the $n$-th chain and $\left(\boldsymbol\theta'_n,\ell'_n,\boldsymbol\tau'_n\right)$ denotes the candidate state. Moreover, we use the notation $\mbox{P}_{\mbox{prop}}\left(x \rightarrow y\right)$ to denote the probability of proposing state $y$ when the current state of the chain is $x$.

\paragraph{Move 1} This move updates the number of change-points, while keeping the mean parameters constant. We introduce two move types which propose to update the total number of change-points by 1. These move-types are complementary to each other: addition/deletion of a change-point. In the following, $\{\boldsymbol\tau_n\cup t\}$ denotes the resulting ordered set when a new change-point $t$ is added 
to the current configuration $\boldsymbol\tau_n$. In a similar fashion, $\{\boldsymbol\tau_n\setminus t\}$ denotes the remaining set when a specific member $t$ of $\boldsymbol\tau_n$ is removed from the c
urrent configuration.

At a given state consisting of $\ell_n$ change-points, we propose addition/deletion with probabilities $p_a(\ell_n)$ and $p_d(\ell_n) = 1 - p_a
(\ell_n)$, respectively. The addition probabilities are defined as
\begin{equation}
p_a(\ell_n) = \begin{cases}
1, & \quad \ell_n = 0\\
1/2, & \quad 1\leqslant \ell_n \leqslant L - 1\\
0, & \quad \ell_n = L,
\end{cases}
\end{equation}
where $L$ denotes the maximum number of change-points ($L\leqslant T-2$). In case of addition, we propose to add a randomly drawn change-point between two successive ones. The probability of proposing the addition of change-point $t_*$ such that $\tau_{n;j-1} < t_* < \tau_{nj}$; for some $j = 1,\ldots,\ell_n+1$, is equal to $p_a(\ell_n)\frac{1}{\tau_{nj} - \tau_{n;j-1} - 1}\frac{1}{\ell_n+1}$. In case that $\tau_{nj} - \tau_{n;j-1} = 1$ the proposed move is immediately rejected. In the reverse move, a the previously added change-point is selected with probability $p_d(\ell_n+1)\frac{1}{\ell_n+1}$ and is deleted from $\boldsymbol\tau_n\cup t_*$. 
Thus, the acceptace probability for an addition move is equal to
\begin{equation}\label{eq:m3a}
\alpha_{a}(\ell_n,\boldsymbol\theta_n, \boldsymbol\tau_n,\boldsymbol\tau'_n) := 
\frac{f\left(\boldsymbol x_n|\boldsymbol\theta_{\{\boldsymbol\tau_n\cup t_*\}}, \boldsymbol \sigma^2_n\right)f\left(\{\boldsymbol\tau_n\cup t_*\}|\ell_n+1\right)(1-p_a(\ell_n+1)) }{f\left(\boldsymbol x_n|\boldsymbol\theta_{\boldsymbol\tau_n}, \boldsymbol \sigma^2_n\right)f\left(\boldsymbol\tau_n|\ell_n\right)\frac{p_a(\ell_n)}{\tau_{nj} - \tau_{n;j-1} - 1} },
\end{equation}
In the case of proposing deletion of a change-point $\tau_{nj}$, the corresponding acceptance ratio term is equal to
\begin{equation}\label{eq:m3b}
\alpha_{d}(\ell_n,\boldsymbol\theta_n,\boldsymbol\tau_n,\boldsymbol\tau'_n) = \frac{1}{\alpha_{a}\left(\ell_n - 1,\boldsymbol\theta_n,\{\boldsymbol\tau_n\setminus \tau_{nj}\}, \boldsymbol\tau_n\right)}.
\end{equation}
At this point we underline that using an overfitted set of model parameters (one mean $\theta_{nt}$ per time-point $t=1,\ldots,T$) allows us to use the standard Metropolis-Hastings ratio for proposing additions/deletions of change-points. This would not be true if the number of mean parameters was defined conditionally on $\ell_n$: in such  a case the Reversible Jump algorithm or integration of mean parameters is required.

\paragraph{Move 2} In this move the update of mean parameters $\boldsymbol{\theta}$ is proposed, while all other parameters remain unchanged. For this purpose a random walk centered at the current values of the chain is used. For subject $n = 1,\ldots,N$, let $\theta_{nt}$  denote the  current value of the mean parameters at time-point $t = 1,\ldots,T$. Then a new state is proposed according to 
\begin{equation*}
\theta'_{nt} \sim \mathcal N\left(\theta_{nt},c \sigma^2_{nt}\right),
\end{equation*}
independent for all $t$ and $n$, for some constant $c>0$. Recall that $\sigma^2_{nt}$ denotes the variance of the random sample $(X_{nt1},\ldots,X_{ntR})$ which is assumed known. Note that the ratio of the proposal distribution for the transitions $\boldsymbol\theta_n\rightarrow\boldsymbol\theta'_n$ and $\boldsymbol\theta'_n\rightarrow\boldsymbol\theta_n$ is 1. Thus, the Metropolis-Hastings acceptance ratio \eqref{eq:mh} simplifies to 
\begin{equation}\label{eq:m4}
\alpha_4(\boldsymbol\tau_n, \boldsymbol\theta_n,\boldsymbol\theta'_n) =  
\frac{f\left(\boldsymbol x_n|\boldsymbol\theta'_{\boldsymbol\tau_n}, \boldsymbol \sigma^2_n\right)f\left(\boldsymbol\theta'_n|\boldsymbol\sigma^2_{n}\right) }{f\left(\boldsymbol x_n|\boldsymbol\theta_{\boldsymbol\tau_n}, \boldsymbol \sigma^2_n\right)f\left(\boldsymbol\theta_n|\boldsymbol\sigma^2_{n}\right) },
\end{equation}

\paragraph{Move 3.a} The candidate state is generated by using a proposal distribution which will jointly update the change-points $\boldsymbol\tau_n$, while the total number of change-points $\ell_n$ and mean parameters $\boldsymbol\theta_n$ are kept constant. Let $\varepsilon = (\varepsilon_1, \ldots, \varepsilon_{\ell_n})$ and $$\varepsilon_i\sim g(\cdot;-d_1,d_1),$$ where $d_1>0$ denotes a pre-specified positive integer and $g(\cdot;-d_1,d_1)$ denotes the discrete uniform distribution over $\{-d_1,-d_1+1,\ldots,d_1-1,d_1\}$. Then, the proposed state is generated as $$\tau'_{ni} = \tau_{ni} + \varepsilon_i$$ independently for $i = 1,\ldots,\ell_n$, while $\ell'_n = \ell_n$ as well as $\boldsymbol\theta'_n = \boldsymbol\theta_n$. In this case, the proposal ratio in Equation \eqref{eq:mh} is equal to 1 so the acceptance ratio is written as the posterior probability ratio.

\begin{equation}\label{eq:m2}
\alpha_1(\ell_n, \boldsymbol\theta_n, \boldsymbol\tau_n,\boldsymbol\tau'_n) := 
\frac{f\left(\boldsymbol x_n|\boldsymbol\theta_{\boldsymbol\tau'_n}, \boldsymbol \sigma^2_n\right)f\left(\boldsymbol\tau'_n|\ell_n\right) }{f\left(\boldsymbol x_n|\boldsymbol\theta_{\boldsymbol\tau_n}, \boldsymbol \sigma^2_n\right)f\left(\boldsymbol\tau_n|\ell_n\right) },
\end{equation}
where $\boldsymbol\theta_{\boldsymbol\tau_n}$ as in \eqref{eq:thetatau}. A small value for $d_1$ will be capable of achieving optimal acceptance rates. Thus, this move is oriented towards the local exploration of the posterior surface, given the current state. 

\paragraph{Move 3.b} This is a similar proposal to Move 3.b, but instead of proposing the simultaneous update of all cut-points, just one entry is modified and the rest remain the same. As in Move 3.a, both the number of change-points as well as the values of mean parameters remain the same. Thus, let $i^{*}$ denote a randomly drawn index from the set $\{1,\ldots,\ell_n\}$ and $$\varepsilon\sim g(\cdot;-d_2,d_2),$$ where $d_2>0$ denotes a pre-specified positive integer. The proposed state is generated as 
\begin{equation}
\tau'_{ni}=\begin{cases}
 \tau_{ni},&\quad i\neq i^{*}\\
  \tau_{ni}+\varepsilon,&\quad i = i^{*}.
\end{cases}
\end{equation}
The Metropolis-Hastings acceptance probability simplifies to Equation \eqref{eq:m2}. In this case, a sufficiently large value for $d_2$ will propose moves that are more likely to be accepted compared to Move 3.a, since only one entry is changed. Thus, move 3.b will be used as complementary to move 3.a, in order to facilitate the ability of escaping from local modes of the posterior distribution. 

\paragraph{Move 4} Let $\boldsymbol\theta_{n[-\boldsymbol\tau_n]}$ denote the mean of those time-points that do not correspond to change-points for time series $n$. In this case it can be easily seen that the full conditional distribution of $(\boldsymbol\theta_{n[-\boldsymbol\tau_n]}|\boldsymbol\tau_n, \boldsymbol x_n, \ell_n)$ is the prior distribution in Equation \eqref{eq:thetaPrior}. Hence, a draw from the prior distribution will perfom a Gibbs update to  $\boldsymbol\theta_{n[-\boldsymbol\tau_n]}$.

Note that both moves 3.a and 3.b are able to propose states that have zero prior probability in Equation \eqref{eq:prior1}. Although this is not a frequent event, in this case the proposed state is immediately rejected, since the prior probability ratio is equal to zero.

Recall that the previous MCMC steps are defined conditionally on $\boldsymbol\sigma^2$. The next subsection deals with the case where the variance is estimated at a pre-processing stage and plugged into the previously described MCMC sampler. For this purpose, we consider that the variance can be either shared between different time series or not and two estimators are derived. We assume that we have enough data in order to obtain a robust estimate of variance, which is the case in our application. The more general case where the variance is treated as a random variable is discussed in subsection \ref{sec:unknownVar}. In this case, the variance is updated by the full conditional distribution using a Gibbs sampling step.

\subsection{Variance estimation at a pre-processing stage}\label{sec:variance}

In this case the variance is considered known and in practice it should be estimated  at a pre-processing stage. We use the posterior mean arising from a multivariate normal--inverse gamma model as a point estimate. For this purpose we ignore the piecewise linear parameterization of the mean function and use the same likelihood as in Equation \eqref{eq:model0} and the same prior assumptions for $\theta_{nt}$ as in Equation \eqref{eq:thetaPrior}. 

We will assume two parameterizations: the full model where the variances are a-priori distributed as:
\begin{eqnarray}\label{eq:freeVar}
\sigma^{2}_{nt} &\sim& \mathcal{IG}\left(\alpha_0,\beta_0\right), 
\end{eqnarray}
independent for $n=1,\ldots,N$; $t=1,\ldots,T$, where $\mathcal{IG}(\alpha,\beta)$ denotes the inverse Gamma distribution. The second model parameterization imposes the restriction of common variance across different time series and replicates, that is, 
\begin{equation}\label{eq:sameVar}
\sigma^2_{nt} = \sigma^2_t\quad n = 1,\ldots,N.   
\end{equation}
Under \eqref{eq:sameVar}, a-priori it is assumed that
\begin{eqnarray*}
\sigma^{2}_{t} &\sim& \mathcal{IG}\left(\alpha_0,\beta_0\right), 
\end{eqnarray*}
independent for  $t=1,\ldots,T$. The quantities $\alpha_0>0$ and $\beta_0>0$ correspond to fixed hyper-parameters. 

Let us define the function
\begin{eqnarray*}
\widehat{\beta}_{nt} & = & \frac{R\nu_0\mu_{0t}^{2} + (R+\nu_0)\sum\limits_{r=1}^{R}x^2_{ntr}-\left(\sum\limits_{r = 1}^{R}x_{ntr}\right)^2-2\nu_0\mu_{0t}\sum\limits_{r=1}^{R}x_{ntr}}{2(R+\nu_0)},
\end{eqnarray*}
for $t = 1,\ldots,T$; $n = 1,\ldots,N$. It easily follows that under \eqref{eq:freeVar}: $$\sigma^2_{nt}|\boldsymbol x\sim \mathcal{IG}\left(\alpha_0 + R/2,\beta_0 + \beta_{nt}\right),$$ independent for $n=1,\ldots,N$; $t=1,\ldots,N$. In the case of the restricted parameterization in \eqref{eq:sameVar}, the corresponding posterior distribution is 
$$\sigma^2_t|\boldsymbol x\sim \mathcal{IG}\left(\alpha_0 + \frac{NR}{2},\beta_0 + \sum_{n = 1}^{N}\widehat{\beta}_{nt}\right),$$ 
independent for $t = 1,\ldots,T$. Then, we use the posterior means as the plug-in point estimates of the variance per time-point, that is,
\begin{eqnarray}
\label{eq:sigmaEstimate0}
\mathbb E\left(\sigma^2_{nt}|\boldsymbol x\right) &=& \frac{\beta_0+\widehat{\beta}_{nt}}{\alpha_0 + \frac{R}{2} - 1},\quad t= 1,\ldots, T; n = 1,\ldots,N\\
\label{eq:sigmaEstimate}
\mathbb E\left(\sigma^2_t|\boldsymbol x\right) &=& \frac{\beta_0+\sum_{n=1}^{N}\widehat{\beta}_{nt}}{\alpha_0 + \frac{NR}{2} - 1},\quad t= 1,\ldots, T,
\end{eqnarray}
provided that $\alpha_0 + R/2 > 1$ so that both posterior expectations exist. The default values we use for the constants in Equations \eqref{eq:sigmaEstimate0} and \eqref{eq:sigmaEstimate} are $\alpha_0 = 1$ and $\beta_0 = 1$, but we also report some prior sensitivity checks in the simulation section. We will use the labels ``s1'' and ``s2'' to refer to the MCMC sampler using the plug-in estimates \eqref{eq:sigmaEstimate0} and \eqref{eq:sigmaEstimate}, respectively.

\subsection{Updating the variance}\label{sec:unknownVar}

Here we discuss the case where the variance in Equation \eqref{eq:model0} is unknown. In the case where all variances are unrestricted (Equation \eqref{eq:freeVar}), the full conditional distributions are:
\begin{equation}
\sigma^2_{nt}|\boldsymbol{x}, \boldsymbol{\theta}, \boldsymbol\tau  \sim \mathcal{IG}\left[\frac{R+1}{2}+\alpha_0,\frac{1}{2}\sum_{r=1}^{R}\{x_{ntr}-\mu(t;\boldsymbol\theta_n,\boldsymbol\tau_n)\}^2+\frac{\nu_0}{2}(\theta_{nt}-\mu_{0t})^2+\beta_0\right],
\end{equation}
independent for $n=1,\ldots,N$, $t =1,\ldots,T$. Note that in this case, the full conditional distribution depends on the piecewise linear mean function $\mu(t;\theta_n,\tau_n)$.  Under restriction \eqref{eq:sameVar}, it follows that:
\begin{equation}
\sigma^2_t|\boldsymbol{x}, \boldsymbol{\theta}, \boldsymbol\tau \sim \mathcal{IG}\left[
\frac{N(R+1)}{2}+\alpha_0,
\frac{1}{2}\sum_{n=1}^{N}\sum_{r=1}^{R}\{x_{ntr}-\mu(t;\boldsymbol\theta_n,\boldsymbol\tau_n)\}^2+\frac{\nu_0}{2}\sum_{n=1}^{N}(\theta_{nt}-\mu_{0t})^2+\beta_0
\right],
\end{equation}
independent for $t=1,\ldots,T$. 

We will use the labels ``s3'' and ``s4'' to refer to the MCMC sampler using the Gibbs steps \eqref{eq:freeVar} and \eqref{eq:sameVar}, respectively.
Since the full conditional distribution of $\sigma_{nt}$ in Equation \eqref{eq:freeVar} depends only on the quantities $(\boldsymbol x_n, \boldsymbol\theta_n, \boldsymbol\tau_n)$, the joint posterior distribution factorizes over $n = 1,\ldots,N$. Thus, we can split the MCMC sampling into $N$ independent samplers, as also done to the case where the variance is fixed. However, this does not hold for Equation \eqref{eq:sameVar} where the state of the parameters $\mu(t;\boldsymbol\theta_n,\boldsymbol\tau_n)$ of all time-series ($n=1,\ldots,N$) is required. This imposes a large computational burden since $N$ can be quite large. Therefore, in the examples of the next section we have only applied the first three MCMC samplers (s1, s2 and s3).

\section{Results}\label{sec:results}

\subsection{Simulation study}\label{sec:sim}

\begin{figure}[t]
\begin{center}
\includegraphics[scale = 0.6]{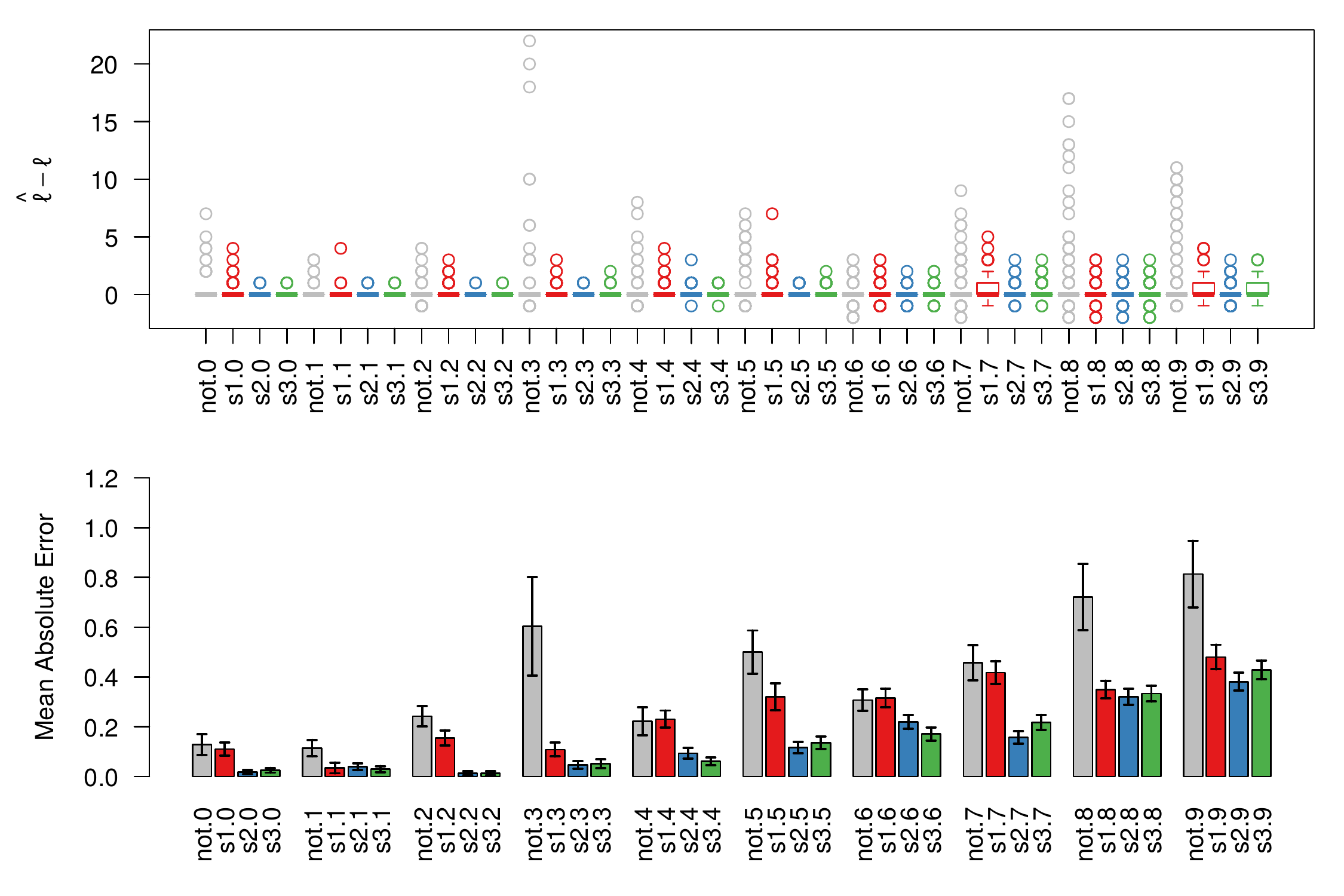}
\end{center}
\caption{Benchmarking the estimation of the number of change-points arising from the narrowest-over-threshold  (``not'') method of \cite{baranowski2016narrowest} and our MCMC sampler with fixed different variance (s1), fixed shared variance (s2) and unknown different variance (s3) per time-series, under the complexity prior distribution. The number after the name of each sampler indicates the true number of change-points and for each value 100 synthetic datasets were simulated. Each time-series consists of three replicates and 1000 time-points.}
\label{fig:threesamplers}
\end{figure}

We considered simulated datasets of length $T = 1000$ time-points consisting of $N = 1000$ independent multivariate observations, while the number of replicates ($R$) is equal to  $R = 3$. For $n = 1,\ldots,N$, the number of change-points $\ell_n$ is drawn uniformly at random from the set $\{0,1,\ldots,9\}$. A detailed description of the simulation mechanism is given in Section B of the Appendix.

All three samplers were applied using the following set of hyper-parameter values: $\alpha_0 = 0.1$, $\beta_0 = 0.1$, $\nu_0 = 0.005$. The prior distribution on the number of change-points corresponds to the complexity prior distribution in \eqref{eq:ellPrior}, with parameters $\alpha = 2$ and $b = 3.72$. The fixed-variance samplers (s1 and s2) are initialized from a state with 1 change-point with parameters randomly generated from the prior distribution. On the other hand, we observed the sampler s3 remains trapped in areas of low posterior probability when using random starting values. In order to deal with this issue, we initialized sampler s3 using a run of sampler s1 (with 30000 iterations). The inference is based on 50000 iterations following a burn-in period of 20000 iterations. 

Our results are benchmarked against the {\tt not} package available in the Comprehensive R Archive Network, which implements the narrowest-over-threshold method of  \cite{baranowski2016narrowest}. The number of change-points is inferred using a strengthened version of the Bayesian Information Criterion \citep{liu1997segmented, fryzlewicz2014wild}. For the problem of detecting changes in the slope, this approach only considers univariate time-series with constant variance, thus, it is not fit for multivariate time-series where the variance varies with time (which is the case in our datasets). In order to apply this method we averaged across the replicates of the time-series. 

Figure \ref{fig:threesamplers} displays the difference of the estimated number of change-points ($\widehat{\ell}$) from the true number ($\ell$), stratified according to the true number of change-points used to generate the data. We conclude that in all cases the estimates are centered in the true value of the number of change-points. At the lower panel of Figure \ref{fig:threesamplers} we calculated the Mean Absolute Error (MAE) of the resulting estimates, with error bars corresponding to the standard error. For smaller number of change-points the MAE is consistently higher for sampler s1. Although the simulation scenario does not assume the same variance per time-series, note that the sampler s2 (which uses the pooled variance estimate in Equation \eqref{eq:sameVar}) gives slightly better results.

We conclude that in many cases the method of \cite{baranowski2016narrowest}  tends to overfit the number of change-points, resulting to worse performance than our method. Clearly, benchmarking against this simpler approach  should not be perceived as a fair comparison between the two methods, but as a means of highlighting the usefulness of our modelling.

\begin{figure}[p]
\centering
\begin{tabular}{c}
\vspace{-3ex}
\includegraphics[scale=0.47]{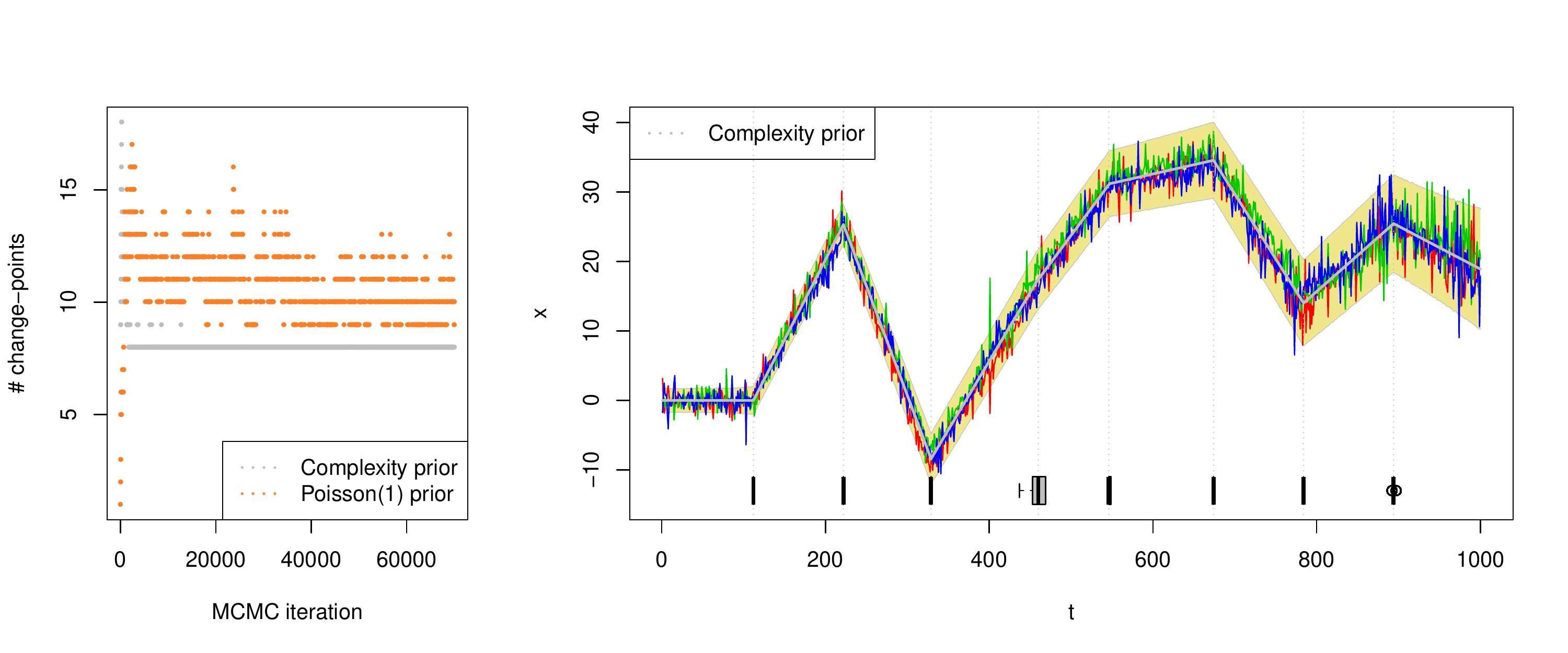}\\
\vspace{-2ex}
\includegraphics[scale=0.47]{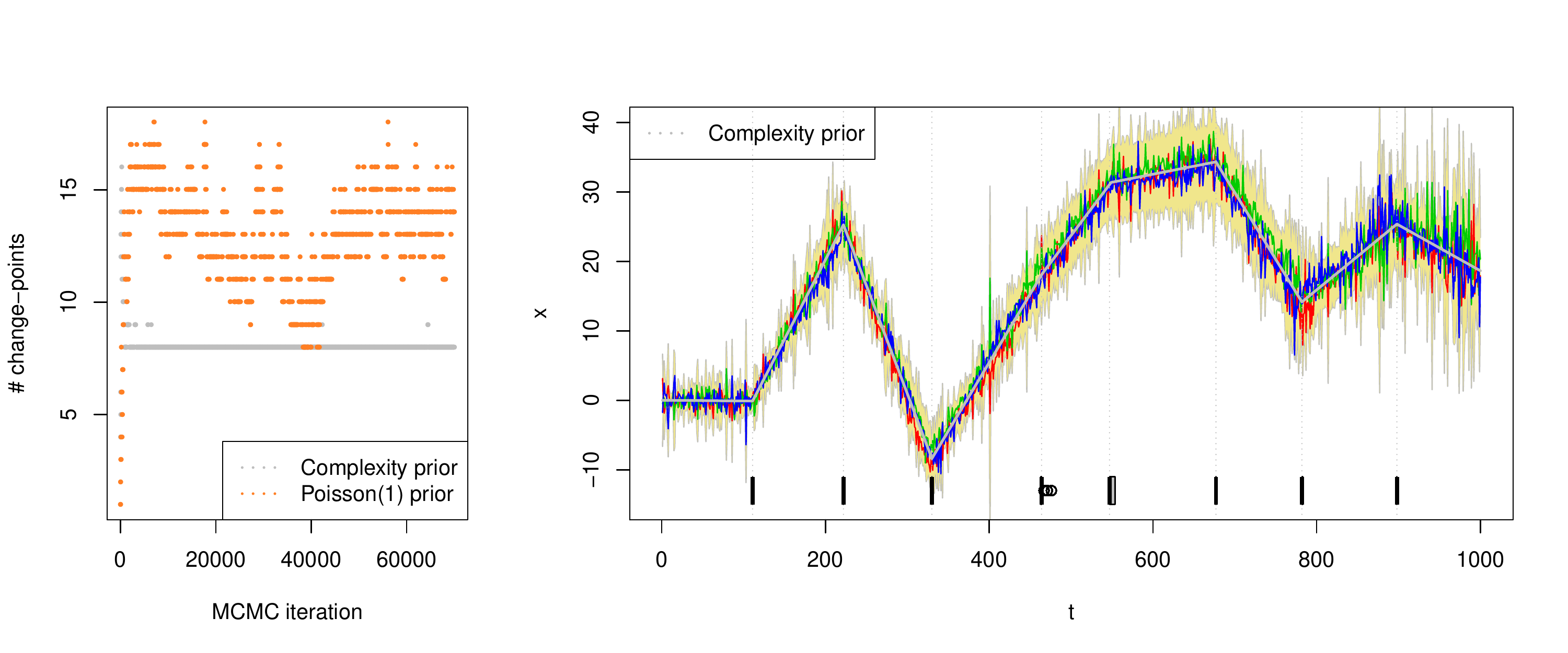}
\end{tabular}
\caption{Example of a simulated time-series with 8 change-points. Replicates are shown in red, blue and green color. Left: MCMC trace of the sampled number of change-points considering both the complexity and a truncated Poisson(1) prior distributions (every 20th iteration is displayed). Right: output of the MCMC sampler conditionally on the estimated MAP of changepoints (which is equal to 8) according to the complexity prior. The upper and lower panels correspond to the MCMC sampler using the variance estimates arising from Equations \eqref{eq:sigmaEstimate} (sampler s2) and \eqref{eq:sigmaEstimate0} (sampler s1), respectively. The gray lines correspond to the posterior mean estimates of the piecewise linear mean function and the boxplots display the posterior marginal distribution of each change-point. Vertical dotted lines correspond to the central line of each boxplot (median). The coloured outer regions correspond to two estimated standard deviations from the mean.}
\label{fig:simExample9}
\end{figure}

Figure \ref{fig:simExample9} displays the output of the MCMC sampler s1 (lower 
panel) and s2 (upper panel) for a time-series where the true number of change-points equals to 8. The output of sampler s3 is almost identical to the lower panel of Figure \ref{fig:simExample9}.  The posterior distribution of the number of change-points is shown at the left panels, considering both the complexity prior distribution (gray trace) as well as a Poisson(1) distribution truncated on the set $\{0,1,\ldots,30\}$. Observe that the first choice quickly converges to a state with 8 change-points (that is, the true number). This is not the case under the Poisson prior distribution, which supports larger values than the true number of change-points. This behaviour is typical in any other simulated dataset we tried, therefore, all results in the remaining sections are based on the complexity prior.  The right panels display the posterior distribution of change-point locations (under the complexity prior) and it evident that the method is able to accurately infer all change-point locations. 

Additional results based on synthetic data are provided in Section B of the Appendix, including prior sensitivity checks as well as alternative simulation scenarios. Further comparisons against {\tt not} are provided in Section \ref{Appsec:not} of the Appendix.

\subsection{Phase detection in parallel time-series analysis of fungal growth}\label{sec:growth}

The filamentous fungal pathogen \textit{Aspergillus fumigatus} is a major pathogen of the human lung causing more deaths per annum than tuberculosis or malaria \citep{brown2012hidden}. A time series study of fungal growth was performed in liquid culture by analysing, in parallel, the growth characteristics of 411 independent transcription factor gene deletion mutants. The mutant strains were cultivated in a microtiter plate containing \SI{200}{\micro\liter} of a fungal culture medium and incubated at \SI{37}{\celsius}. Optical density (at \SI{600}{\nano\meter}) was measured at 10 minute intervals for a total period of 48 hours. The growth analysis was performed on three separate occasions.

The observed data consists of $N\times R \times T$ growth levels for $R = 3$ replicates of $N = 411$ objects (mutants) measured every 10 minutes for $T = 289$ time-points. Figure \ref{fig:timeCourse} displays the observed time series for four mutants. Visual inspection reveals that describing growth with a piecewise linear mean function with an unknown number of segments is a reasonable assumption for the observed data. Regarding the fixed hyper-parameter values, we considered that $\alpha = 2$ (Equation \eqref{eq:ellPrior}) and $\nu_0 = 10^{-1}$ (Equations \eqref{eq:thetaPrior} and \eqref{eq:sigmaEstimate}). After estimating the variance per time-point using the estimator in \eqref{eq:sigmaEstimate}, the MCMC sampler s2 ran for $m = 50000$ iterations, following a burn-in period of $20000$. 

\begin{figure}
\centering
\includegraphics[scale=0.5]{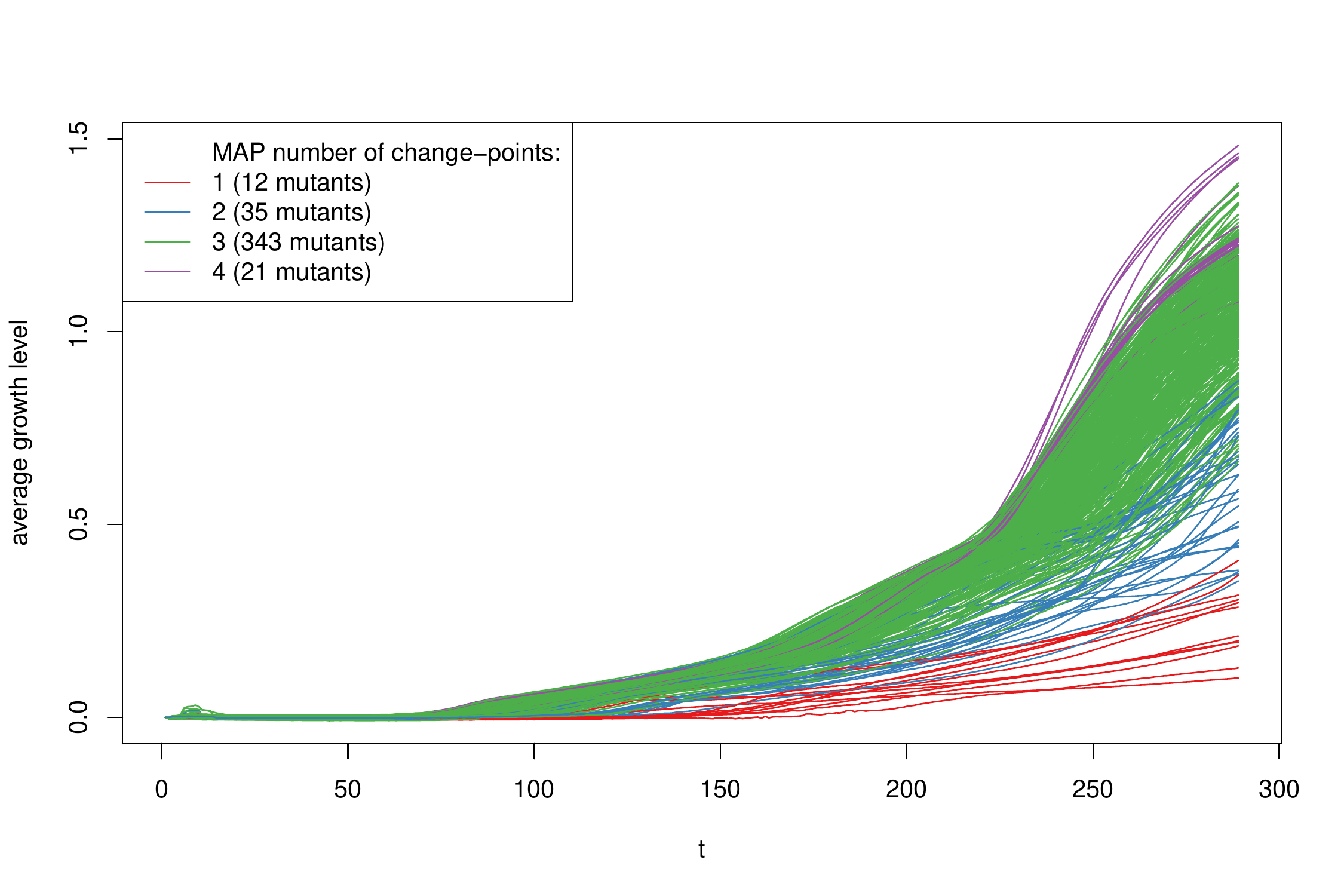}
\caption{Visualization of the growth dataset with respect to the estimated MAP number of phases. For plotting convenience, each curve corresponds to the average growth time series across the three replicates.}
\label{fig:timeSeriesMAP}
\end{figure}

The boxplots in Figure \ref{fig:timeCourse} correspond to the estimate of the marginal posterior distribution of each change-point for specific subset of four mutants, conditionally on the mode of the posterior distribution of the number of change-points. Figure \ref{fig:timeSeriesMAP} displays the averaged profile per mutant (mean of three replicates) coloured according to the most probable number of change-points for each of $N = 411$ subjects. We conclude that the majority of the mutants (343) consist of three growth phases. It is clear that mutants with a smaller number of change-points are also the more slowly growing mutants, which is reasonable since these mutants most likely have not been able to reach the later growth phases in the time of the experiment. In particular the method inferred 35 mutants with only 2 growth phases and slow growth behaviour while 12 mutants have a single phase and very slow growth behaviour. Finally, 21 mutants consist of 4 growth phases and some of them exhibit a faster growth rate at later observation stages ($t > 220$). 

Amongst 12 fungal mutants identified as having a single change-point during growth curve analysis, and therefore exhibiting severely retarded growth kinetics, seven mutants had previously been characterised \citep{lee2016negative, 10.1371, 10.1371/journal.ppat.1005775, bertuzzi2014ph, 10.1371/journal.ppat.1003573, dinamarco2012molecular}. Without exception previously characterised mutants had been reported as  having various morphological defects. The remaining five mutants have, until now, remained uncharacterised and therefore provide promising candidates to investigate further in order to establish their roles in fungal morphogenesis.

Amongst the cohort of known morphogenesis phenotypes correctly identified in our analysis, the transcription factor null mutant \textit{$\Delta$nsdC} (lacking the AFUB\_089440 gene) is defective in acquisition of developmental competence and exhibit dysmorphic spores, rapid germination kinetics, restricted hyphal growth and developmental abnormalities prompting conidiogenesis from inappropriately differentiated hyphae \citep{lee2016negative} 
The sterol-regulatory element binding protein (SREBP) \textit{SrbA} (encoded by AFUB\_018340), required for cell polarity, hypoxia adaptation, and azole drug resistance is critical for normal hyphal branching and cell polarity \citep{10.1371}. The HapB component of the multimeric \textit{A.~fumigatus} CBC transcription factor complex (encoded by AFUB\_030360), which antagonises the role of the SREBP family, including SrbA, is required for normal growth and loss of HapB function results in a severe growth deficit \citep{10.1371/journal.ppat.1005775}. The pH-responsive \textit{A.~fumigatus} transcription factor \textit{$\Delta$pacC} (encoded by AFUB\_037210) is required for normal colonial growth on supplemented solid DMEM medium pH 7.4. In contrast to colonies of wild type isolates, \textit{$\Delta$pacC} mutants exhibit fewer peripheral invasive hyphae and are composed of a denser hyphal network due to a hyperbranching morphology \citep{bertuzzi2014ph}. Null mutants of the MetR transcription factor \citep{10.1371/journal.ppat.1003573} (encoded by AFUB\_063610) demonstrate reduced rates of spore germination and germ tube formation and null mutants of SebA (encoded by AFUB\_066180) demonstrate reduced growth rates under various nutrient limiting conditions \citep{dinamarco2012molecular}.

We have also considered a Poisson(1) prior distribution on the number of change-points. As already demonstrated in Section \ref{sec:sim}, in this case the sampler selects a larger number of change-points which are less interpretable. The reader is referred to Section C of the Appendix.

\section{Discussion}\label{sec:discussion}

A method for inferring the number of change-points in the underlying piecewise linear mean function of replicated time-series has been presented. A crucial characteristic of the model is that each time-point may have its own mean, an assumption which introduces an overfitting number of parameters. The method is able to penalize overfitting models by using an exponentially decreasing prior distribution \citep{castillo2012} on the number of change-points and it was demon
strated that this approach leads to a posterior distribution that can accurately recover the underlying sparse structure of the model. 

We considered that the variance may be fixed or treated as unknown. In the first case we used a plug-in estimate arising from a pre-processing stage of the data. Moreover two different variance parameterizations were introduced, depending on whether the variance of each time-point is shared between different time-series or not. We have also discussed the case where the variance is treated as unknown, updating its state by an extra Gibbs sampling step. According to our simulations, the samplers with fixed variance (s1 and s2) are quite competitive with the sampler s3 which also updates the variance. From a Bayesian point of view, it is preferable to use sampler s3 since all parameters are jointly inferred, however, samplers s1 and s2 are more beneficial from a computational point of view.

There are many interesting extensions of our research.  For example, one could assume more general models between replicates, such as a multivariate normal distribution with full covariance matrix and/or replicate-dependent means, or even models that are not necessarily normal. The core mechanism of the proposed MCMC sampler will be the same in these situations and it would be interesting to investigate whether the method can produce robust results in such settings. In our setup we observed that our sampler does not face any convergence issues and quickly reaches to a state where the number of change-points reflects the underlying structure of the model. In the previously mentioned generalizations however, it might be beneficial to seek ways of improving the mixing and accelerating convergence by e.g.~embedding our sampler to parallel-tempering schemes. Another interesting extension of our research is to explore the usage of data transformations for stabilizing the variance of time series along time. 

In our biological application, we found that all of the slow-growing mutant strains identified by our method, and which had previously been characterised in the literature, were known to play roles in fungal morphogenesis. Further experiments are planned to explore how the growth dynamics of the mutants considered here changes under different environmental conditions. Our simple change-point method provides a useful low-dimensional model of the growth dynamics to explore gene-environment effects on the growth phenotype. 

\section{Software and data availability}
\label{sec5}

Our algorithm is available as an {\tt R} package \citep{beast} at the Comprehensive R Archive Network \citep{rcitation}. Scripts to reproduce real and simulated data analysis are available online at \texttt{https://github.com/\\mqbssppe/growthPhaseMCMC}.

\section*{Acknowledgements}

The authors would like to acknowledge the assistance given by IT services and use of the Computational Shared Facility of the University of Manchester. The suggestions of two anonymous reviewers helped to improve the findings of this study. Research was funded by MRC award MR/M02010X/1. The  authors  have  declared  no  conflict  of interest.\vspace*{-8pt}

\appendix
\renewcommand{\theequation}{\thesection.\arabic{equation}}
\renewcommand\thefigure{\thesection.\arabic{figure}}
\section{Details of the MCMC sampler}

The following pseudo-code summarizes the workflow of the proposed Metropolis--Hastings MCMC sampler for samplers s1 and s2. 

\begin{enumerate}
\item Pre-processing step to estimate the variance according to Equation \eqref{eq:sigmaEstimate0} (for sampler s1) or \eqref{eq:sigmaEstimate} (for sampler s2).
\item For $n = 1,\ldots,N$
\begin{enumerate}
\item Give some initial values $\boldsymbol\ell_n^{(0)},\boldsymbol\theta^{(0)}_n,\boldsymbol\tau^{(0)}_n$
\item For $m = 1,\ldots,M$
\begin{enumerate}
\item \textbf{Move 1:} Propose addition or deletion of a change-point and accept the candidate state $\ell'_n, \boldsymbol\tau'_n$ according to Equation \eqref{eq:m3a} or \eqref{eq:m3b}, respectively. In case of acceptance set $\left(\ell_n^{(m)},\boldsymbol\tau_n^{(m)}\right) = \left(\ell'_n, \boldsymbol\tau'_n\right)$, otherwise $\left(\ell_n^{(m)}, \boldsymbol\tau_n^{(m)}\right) = \left(\ell_n^{(m-1)}, \boldsymbol\tau^{(m-1)}_n\right)$.
\item \textbf{Move 2:} Propose to update the mean parameters and accept the candidate state $\boldsymbol\theta'_n$ according to Equation \eqref{eq:m4}. In case of acceptance set $\boldsymbol\theta_n^{(m)} = \boldsymbol\theta'_n$, otherwise $\boldsymbol\theta_n^{(m)} = \boldsymbol\theta_n^{(m-1)}$.
\item Propose to update the change-points: with probability 0.5 choose \textbf{Move 3.a}, otherwise choose \textbf{Move 3.b}. Accept the candidate state $\boldsymbol\tau'_n$ according to \eqref{eq:m2}. In case of acceptance set $\boldsymbol\tau_n^{(m)} =  \boldsymbol\tau'_n$.
\item \textbf{Move 4:} Update the mean parameters that are not allocated to a change-point by sampling from the prior distribution.
\end{enumerate}
\end{enumerate}
\end{enumerate}

In the case that the variance is unknown,the sampler is augmented by an extra Gibbs sampling step, which will update the variances using the full conditional distributions descibed in Section \ref{sec:unknownVar} of the paper.

In the presented applications we considered that the total number of MCMC iterations is equal to $M = 70000$, while the first 20000 are discarded as burn-in period. The parameters of the MCMC sampler are defines as follows: $c = 0.05$ (Move 2), $d_1 = 1$ (Move 3.a), $d_2 = T/20$ (Move 3.b). The MCMC sampler is initialized from a state with one randomly selected change-point.  The parameters $\boldsymbol\theta_{n}$ are initialized from the mean of the multivariate normal distribution corresponding to the posterior distribution arising from Equation \eqref{eq:thetaPrior} and \eqref{eq:freeVar}, for $n = 1,\ldots,N$. 

\section{Simulation study details}\label{Appsec:sim}

We considered simulated datasets of length $T = 1000$ time-points consisting of $N = 1000$ independent multivariate observations. Three simulated datasets of dimensionality $N\times R\times T$ were genera
ted according to \eqref{eq:model} considering that the number of replicates ($R$) is equal to  $R = 3$ or $6$. For $n = 1,\ldots,N$, the number of change-points $\ell_n$ is drawn uniformly at random
 from the set $\{0,1,\ldots,9\}$. Given $\ell_n>0$, the ``global position'' of the $j$-th change-point is generated as 
\begin{equation}\label{eq:global}
\tau_j = \frac{T}{(\ell_n+1)}j + y_j, \quad j = 1,\ldots,\ell_n
\end{equation}
where $y_j$ denote independent draws from the  Binomial$(100, 0.5)$ distribution, $j = 1,\ldots,\ell_n$; $n = 1,\ldots,N$. According to this scheme note that $\mathbb E(\tau_{j+1} - \tau_{j}) = \frac{T}{(\ell_n+1)}$ and $\mbox{Var}(\tau_{j+1} - \tau_{j}) = 50$. 

\begin{figure}
\begin{center}
\includegraphics[scale = 0.65]{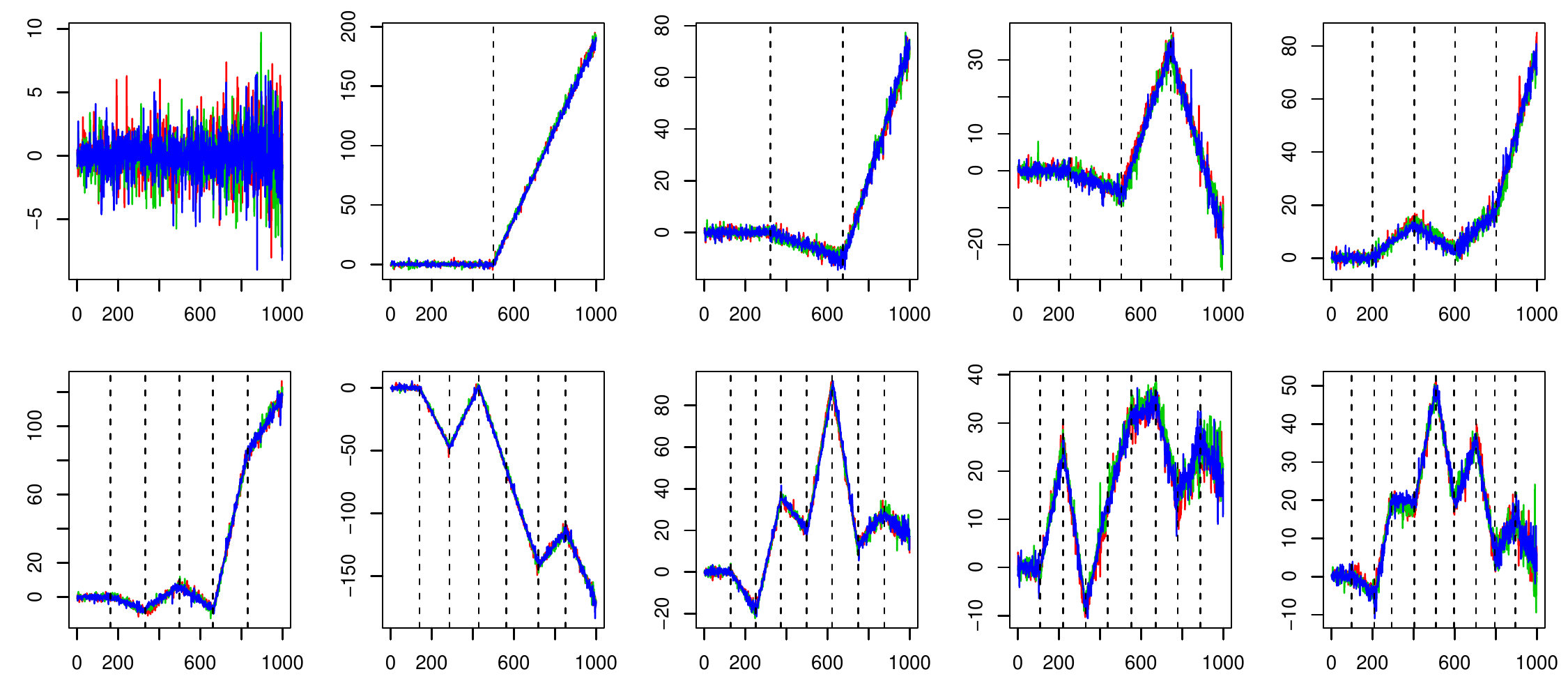}
\end{center}
\caption{An example of 10 subjects from the simulated time-series with $T = 1000$ time-points with 3 replicates (blue, red, green) and number of change-points equal to $0,1,\ldots,9$ (indicated by dashed vertical lines).}
\label{fig:simExamples}
\end{figure}

Recall that our model assumes that the replicates are iid random variables for a given time-point $t$ and time-series $n$. However, we would rather benchmark our method in case that this assumption is not necessarily true. In order to introduce extra noise between replicates we assumed that the true change-point for each replicate has some variation around the time-point in Equation \eqref{eq:global}. Therefore, it was considered that replicate $r$ changes its mean function at time 
\begin{equation}\label{eq:replicateSim}
\tau'_{jr} = \tau_j + dz_{jr}
\end{equation}
where $d\sim\mathcal{DU}\{-1,1\}$ and $z\sim\mathcal P(2)$, independent for $r$, $j$, where $\mathcal P(\cdot)$ denotes the Poisson distribution. The  expression level of each subject $n = 1,\ldots,N$ was initialized at time-point $t=1$ by a baseline mean equal to zero for the first phase. In case that the number of change-points for subject $n$ is positive ($\ell_n >0$), the expression level for each phase was assumed to correspond to a linear function by randomly generating the slope $S_j$ for each consecutive phase according to $S_j = \omega_j|Y_j|$ where $Y_j\sim \mathcal N(0,0.3^2)$, independent for $j = 1,\ldots,\ell_n$. $\omega_j$ denotes a discrete random variable with values in $\{-1,1\}$, distributed according to a Markov process with $\mathbb{P}(\omega_j = 1|\omega_{j-1}=-1) = \mathbb{P}(\omega_j = -1|\omega_{j-1}=1) = p$, where $p = 0.8$.

Each replicate was allowed to have some variation around the endpoints of each phase according to the $\mathcal N(0,\sigma^2_{\mbox{rep}})$ distribution, where 
\begin{equation}\label{eq:sigmaRep}
\sigma^2_{\mbox{rep}} = 1 \ .
\end{equation} 

\subsection{Simulation study 1: different variance per time-series}

Regarding the observation variance $\sigma^{2}_{nt}$ it is  assumed that $$\sigma^2_{nt}\sim\mathcal G\left(1, -\frac{0.9}{T-1}t + \frac{T-0.1}{T-1}\right),$$ independent for $t=1,\ldots,T$ and $n=1,\ldots,N$. Note that $\{\sigma^{2}_{nt};n=1,\ldots,N\}$ is a random sample of size $N$ from a Gamma distribution. In order to estimate the variance per time point and time series both estimators \eqref{eq:sigmaEstimate0} and \eqref{eq:sigmaEstimate} are considered.

Figure \ref{fig:simExamples} illustrates a random subset of $10$ simulated time-series with a number of change-points ranging in $0,1,\ldots,9$. Note that there is strong heterogeneity between time-series
 and that there are instances where replicates deviate from the iid assumption. We will refer to the specific data generation procedure with the term ``noisy scenario''. Note that when $d = 0$ in Equation
 \eqref{eq:replicateSim} and $\sigma^2_{\mbox{rep}} \rightarrow 0$ in Equation \eqref{eq:sigmaRep} the simulation scenario does not introduce any extra noise between replicates and the data is generated b
y the assumptions imposed by the proposed model. Note that the results of this simulation procedure are shown in Figure \ref{fig:threesamplers} of the main paper.

\begin{figure}
\centering
\begin{tabular}{cc}
\includegraphics[scale=0.33]{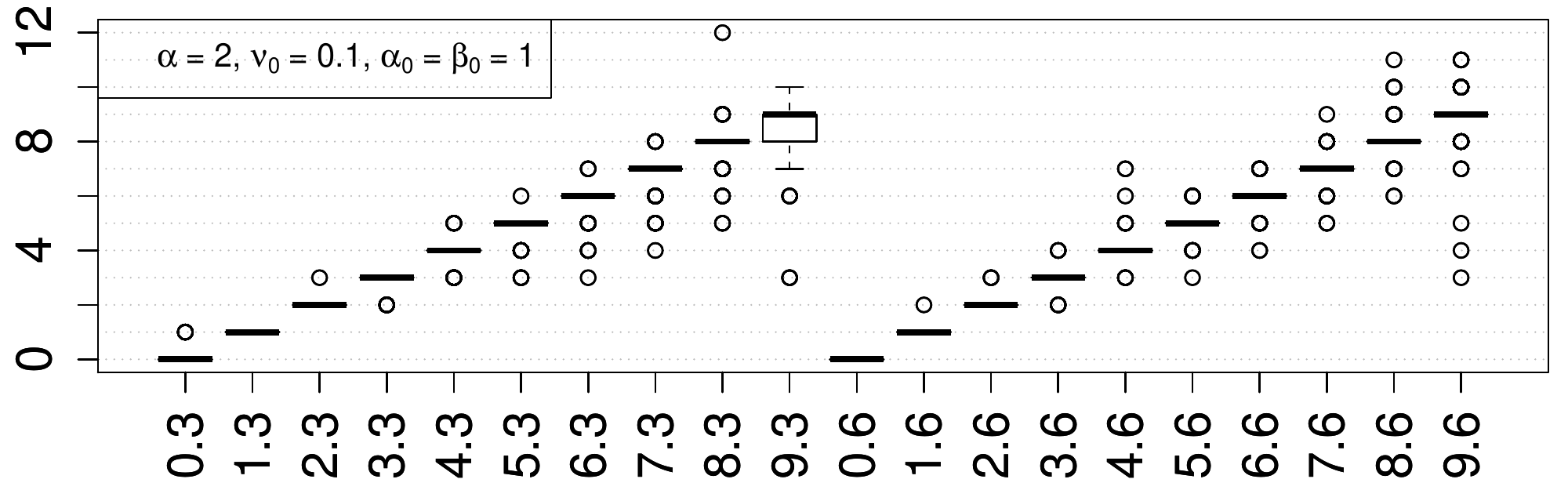}&
\includegraphics[scale=0.33]{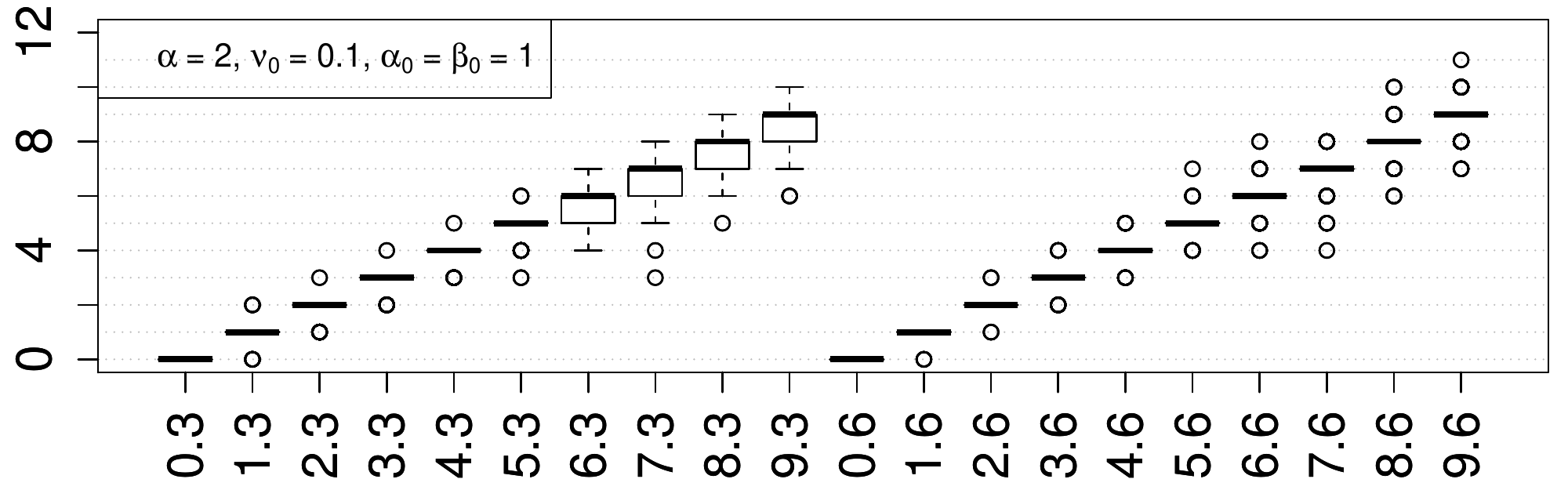}
\\
\includegraphics[scale=0.33]{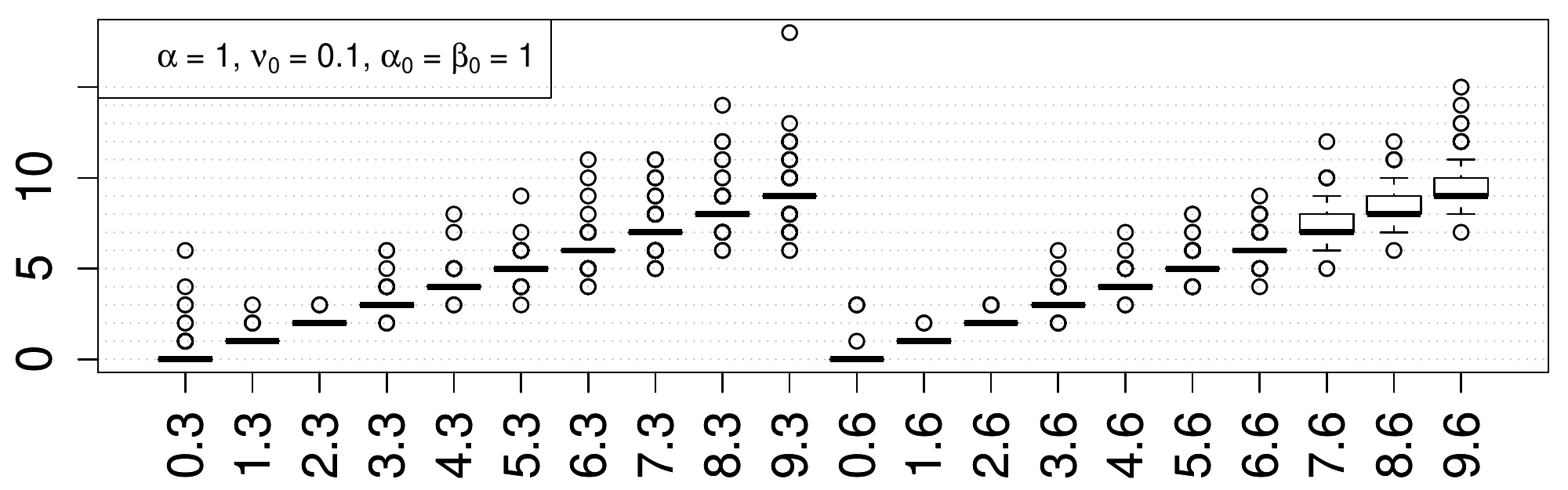}&
\includegraphics[scale=0.33]{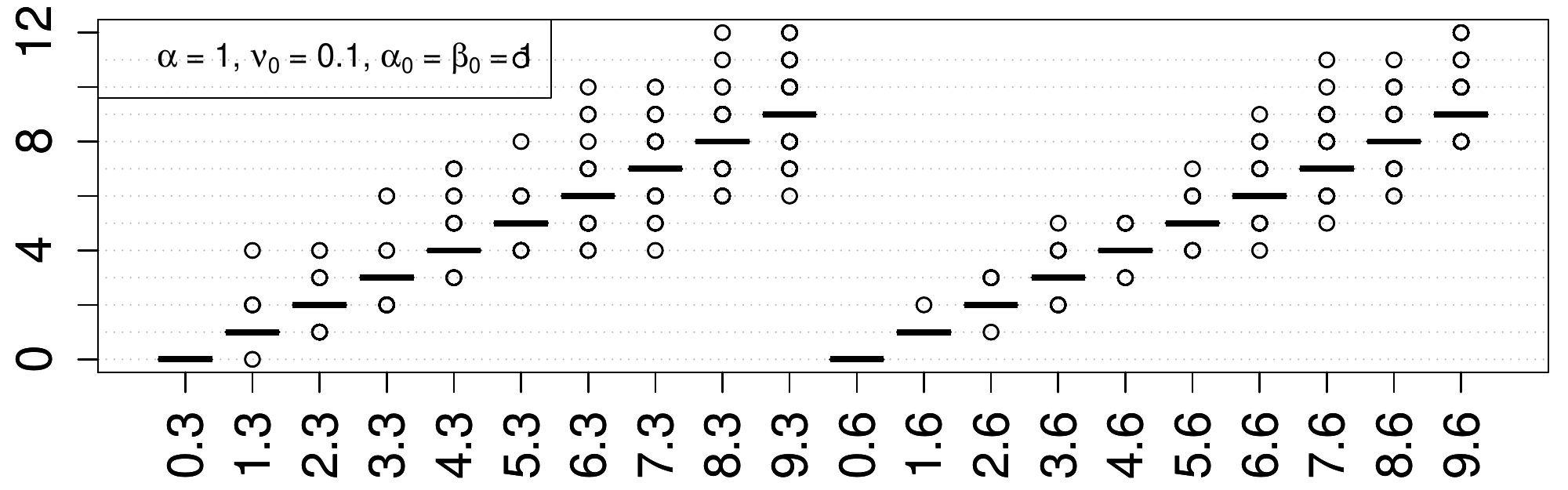}
\\
\includegraphics[scale=0.33]{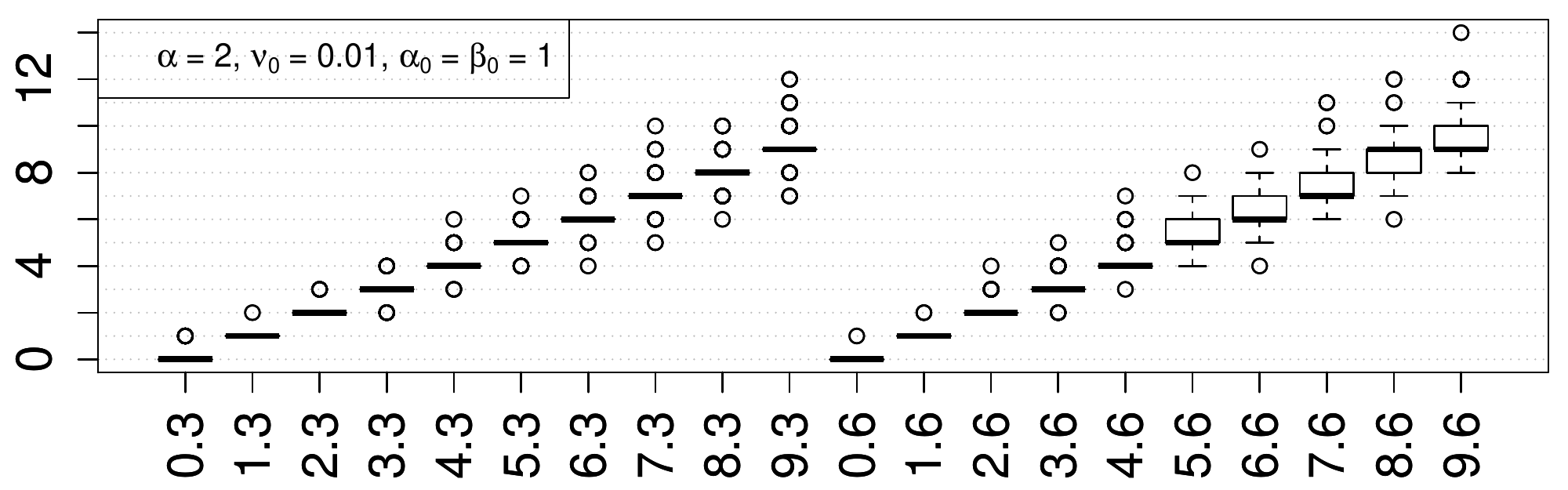}&
\includegraphics[scale=0.33]{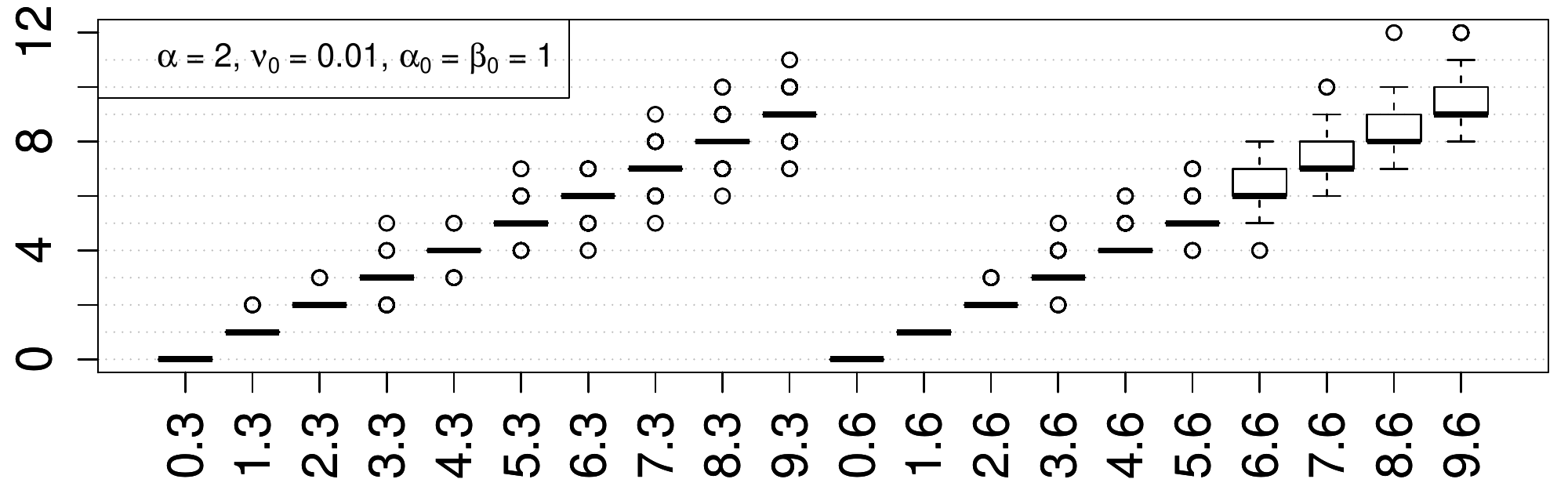}
\\
\includegraphics[scale=0.33]{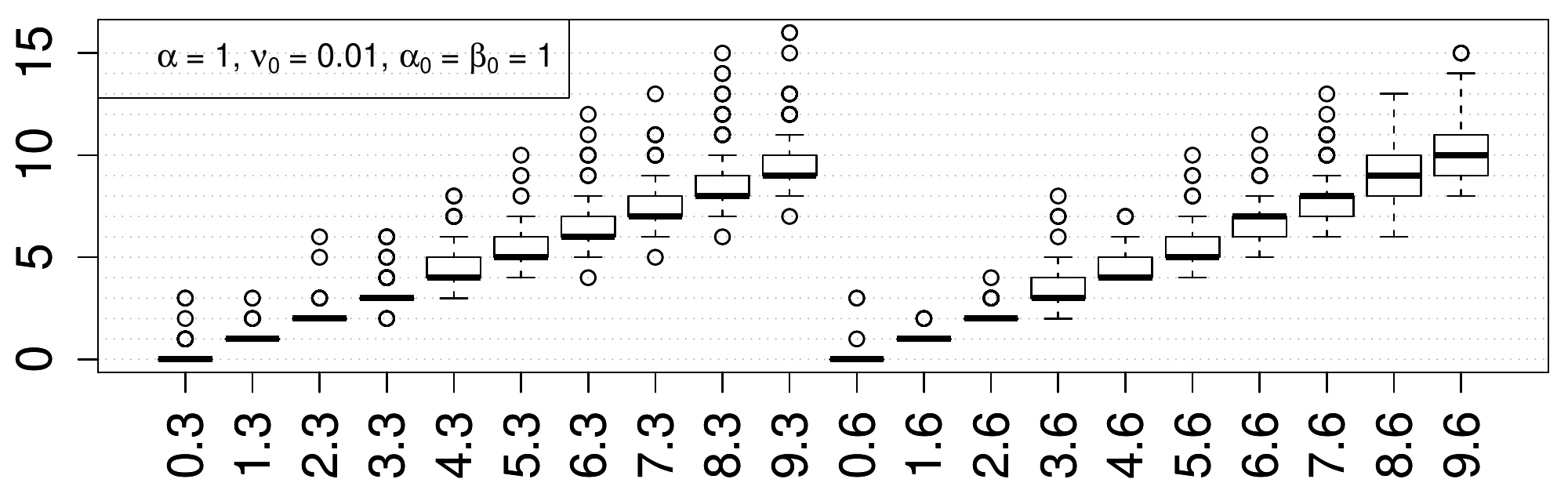}&
\includegraphics[scale=0.33]{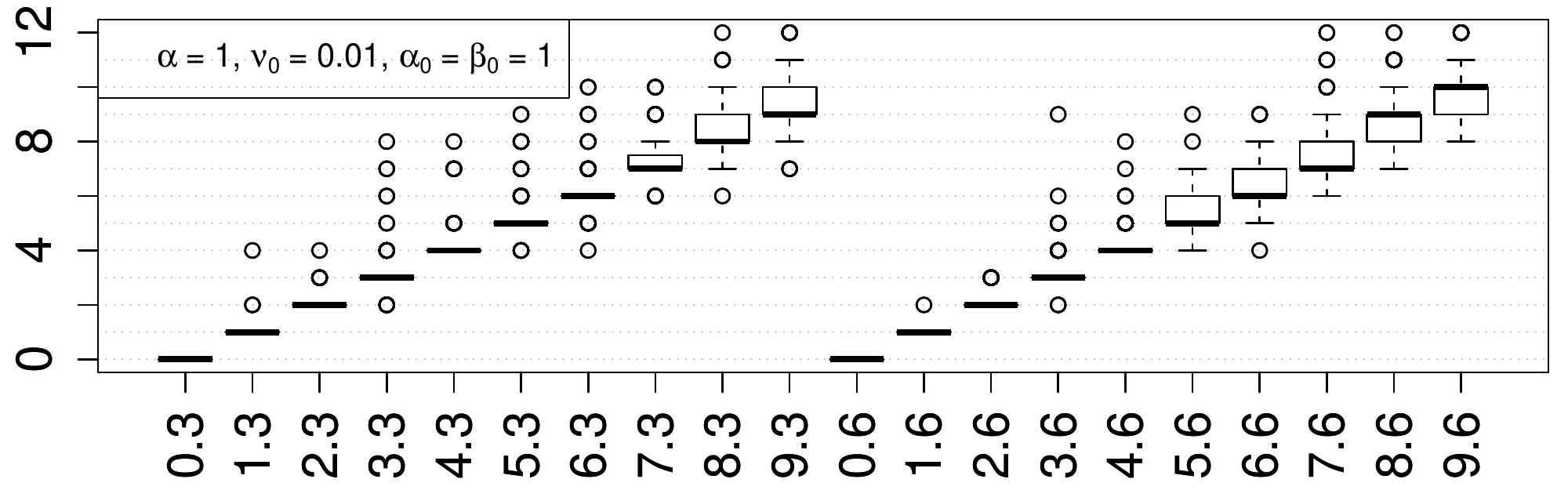}
\\
\includegraphics[scale=0.33]{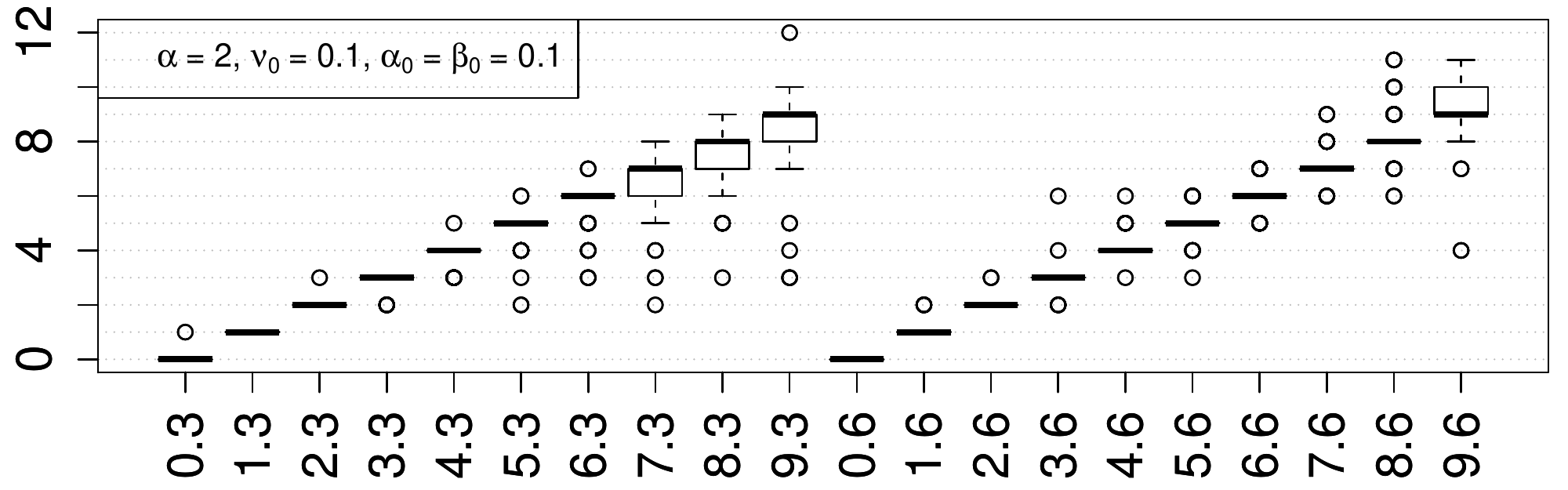}&
\includegraphics[scale=0.33]{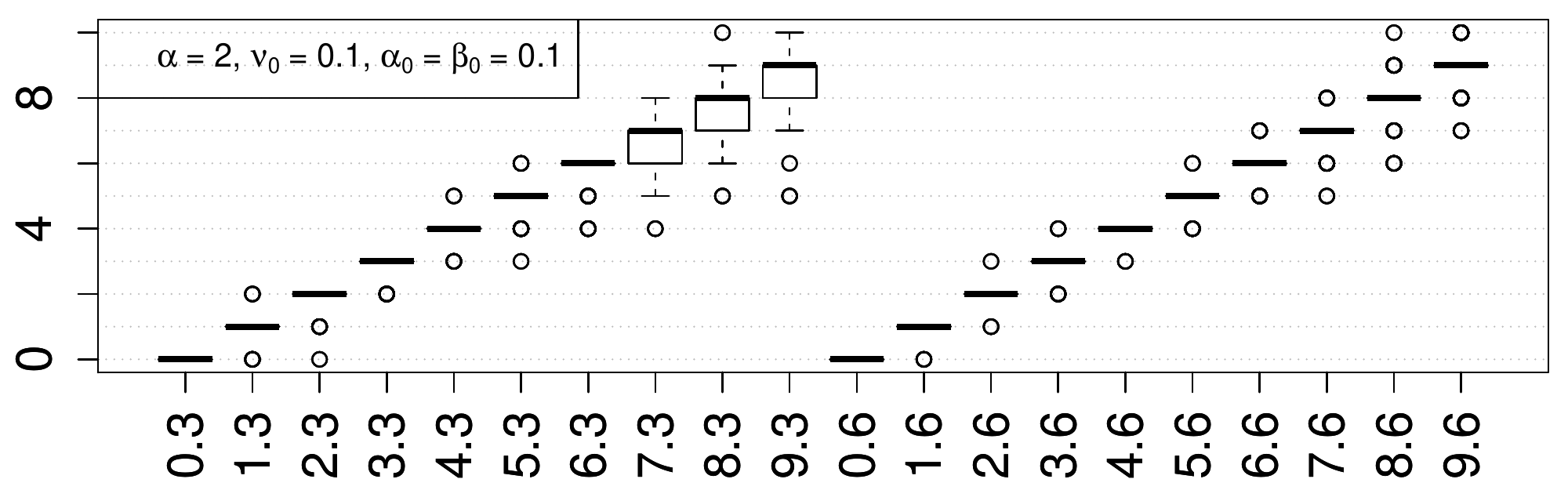}
\\
\includegraphics[scale=0.33]{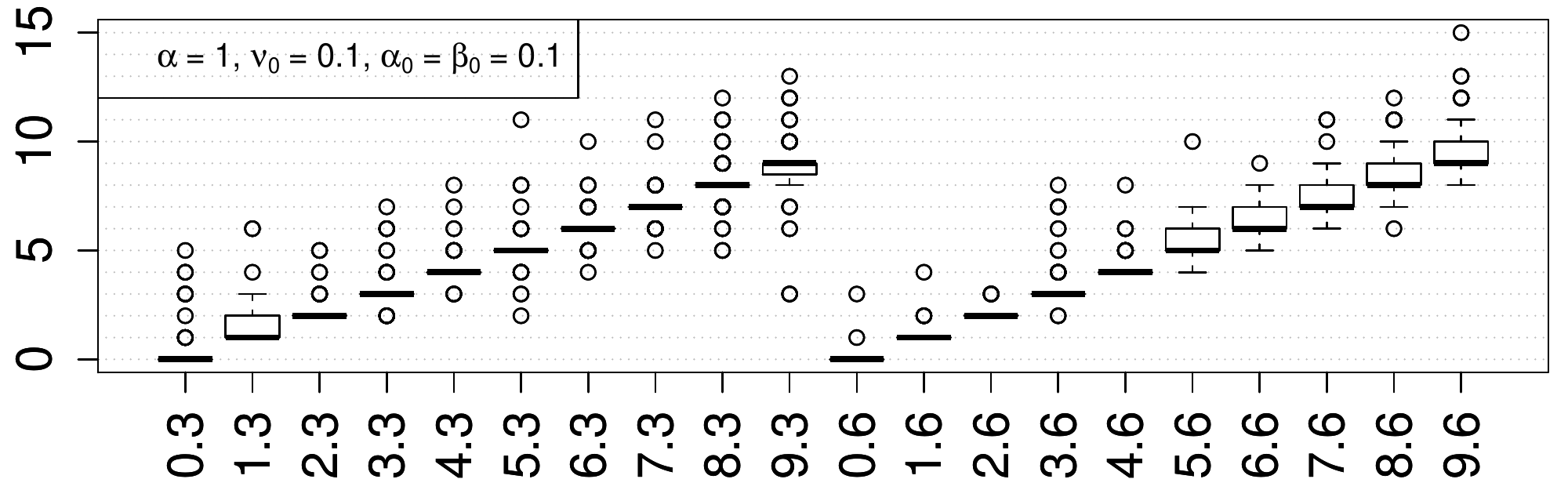}&
\includegraphics[scale=0.33]{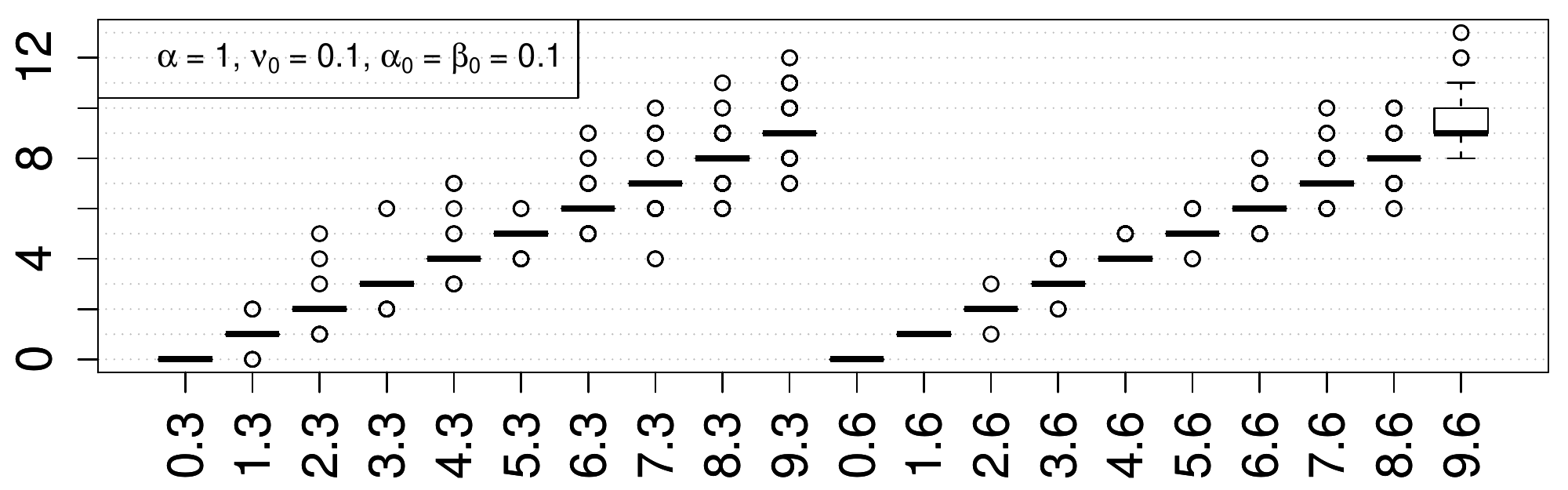}
\\
\includegraphics[scale=0.33]{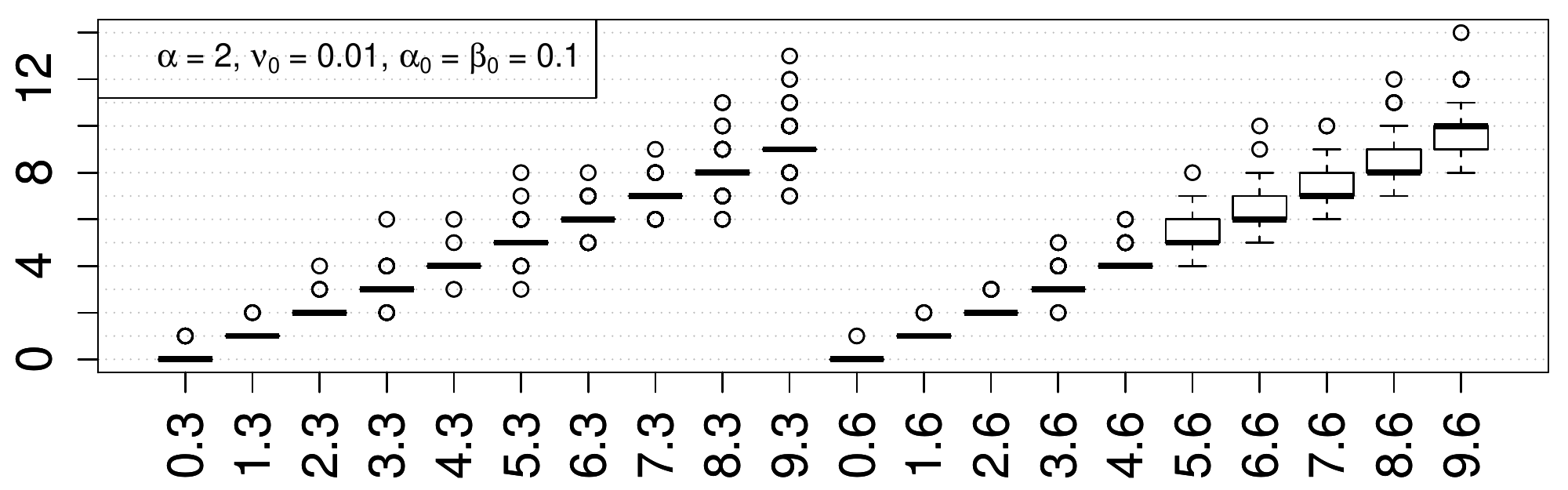}&
\includegraphics[scale=0.33]{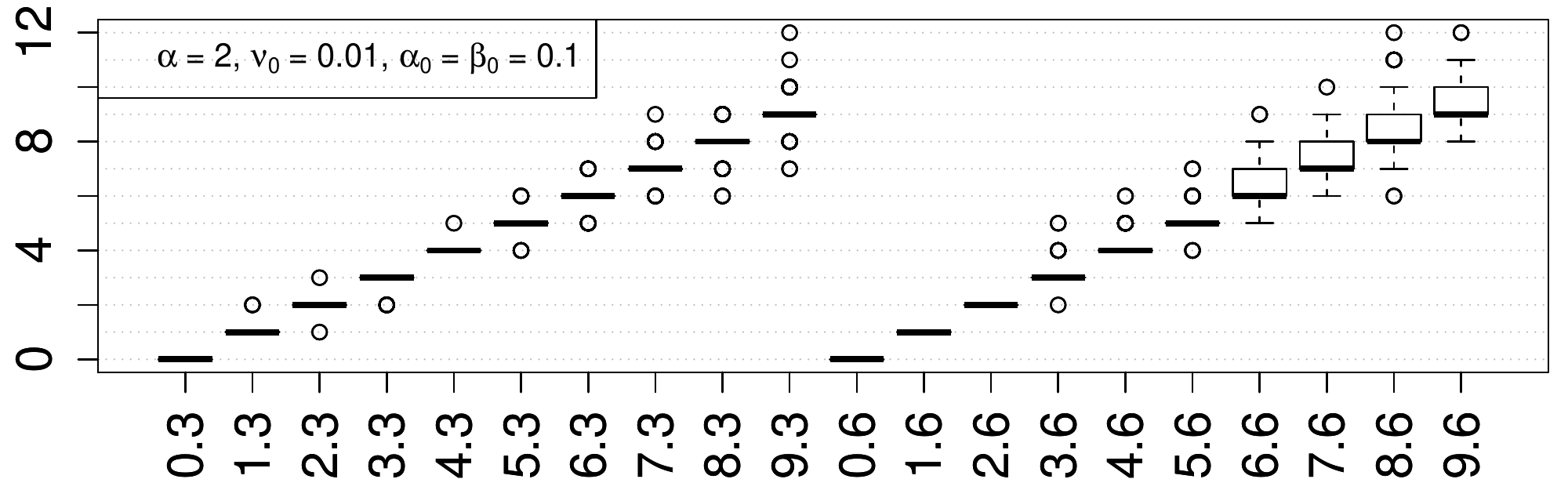}
\\
\includegraphics[scale=0.33]{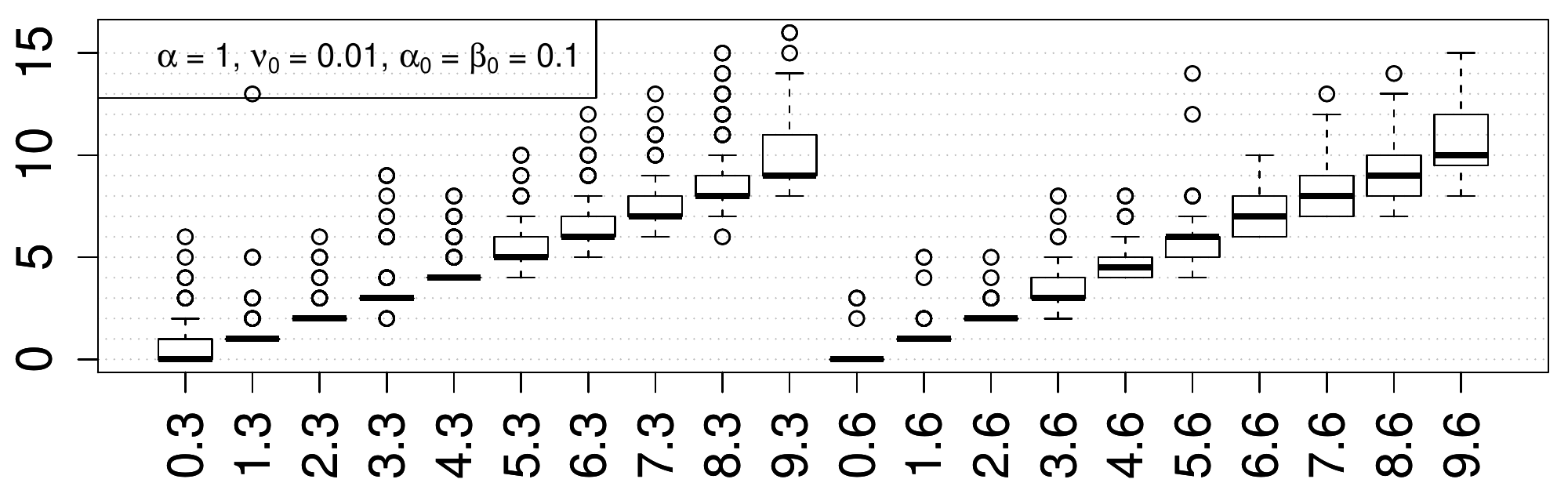}&
\includegraphics[scale=0.33]{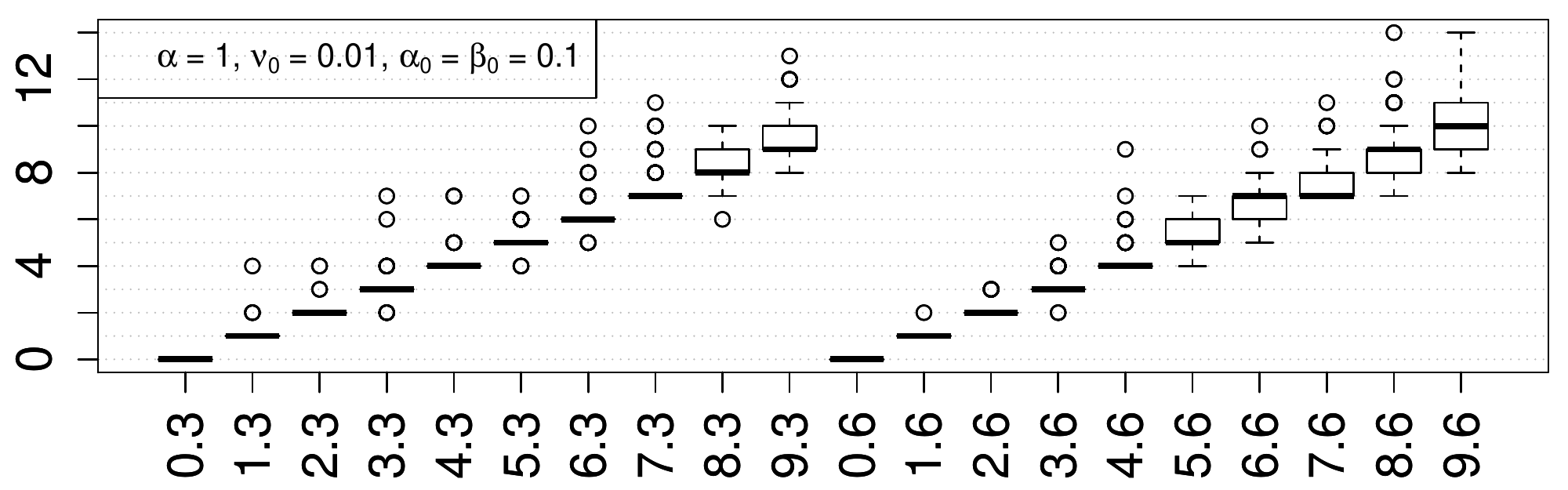}
\\
(a) Using variance estimator \eqref{eq:sigmaEstimate0} & (b) Using variance estimator \eqref{eq:sigmaEstimate} 
\end{tabular}
\caption{Estimation of the number of change-points on synthetic datasets 
generated under different variance per time series. The variance was estimated according to (a): the different variance estimator \eqref{eq:sigmaEstimate0} and (b): the same variance estimator \eqref{eq:sigmaEstimate}. Different combinations of prior parameters $(\alpha, \nu_0,  \alpha_0,\beta_0)$ were used to the MCMC sampler. Each pair of numbers in the horizontal axis displays the true number of change-points (first entry) and number of replicates (second entry).}
\label{fig:change-pointsEstimationDifferentVariance}
\end{figure}

Next we perform some prior sensitivity checks by considering different combinations of the hyper-parameters. Figure \ref{fig:change-pointsEstimationDifferentVariance} shows the selected number of change-points using the approximate MAP estimate from the MCMC sample. Prior-sensitivity checks are performed by considering that $\alpha\in\{1, 2\}$ in Equation \eqref{eq:ellPrior}, $\nu_0\in\{0.01,0.1\}$ in Equation \eqref{eq:thetaPrior} and $(\alpha_0,\beta_0)\in\{(0.1,0.1), (1,1)\}$ in Equations \eqref{eq:sigmaEstimate0}-\eqref{eq:sigmaEstimate}. The results are stratified with respect to the true number of change-points used to generate each time series and we conclude that it is accurately estimated in most cases. Recall that the parameter $\alpha$ controls how fast is the exponential decrease in the prior distribution of the number of change-points, hence larger values of $\alpha$ yield heavier penalties for complex models. This  behaviour is reflected in Figure \ref{fig:change-pointsEstimationDifferentVariance} where we observe that $\alpha = 1$ tends to produce larger MAP estimates than $\alpha = 2$. Observe also that the results are reasonably robust with respect to the parameter $\nu_0$. 

\subsection{Simulation study 2: same variance per time-series}

In this section we replicate the ``noisy'' simulation scenario of Section \ref{Appsec:sim} but now the variance is restricted to be the same between time series.  More specifically, it is  assumed that $\sigma^{2}_{nt} = \sigma^2_t$ for all $n = 1,\ldots,N$ and that $$\sigma^2_{t}\sim\mathcal G\left(1, -\frac{0.9}{T-1}t + \frac{T-0.1}{T-1}\right),$$ independent for $t=1,\ldots,T$. Furthermore, we also consider the ``exact'' simulation scenario where no additional noise is introduced between replicates and expression levels. In order to estimate the variance per time point and time series we used the estimator in  \eqref{eq:sigmaEstimate}. Then, the MCMC sampler was run with the same number of iterations and burn-in period as previously (70000 and 20000, respectively). 

\begin{figure}
\centering
\begin{tabular}{cc}
\includegraphics[scale=0.33]{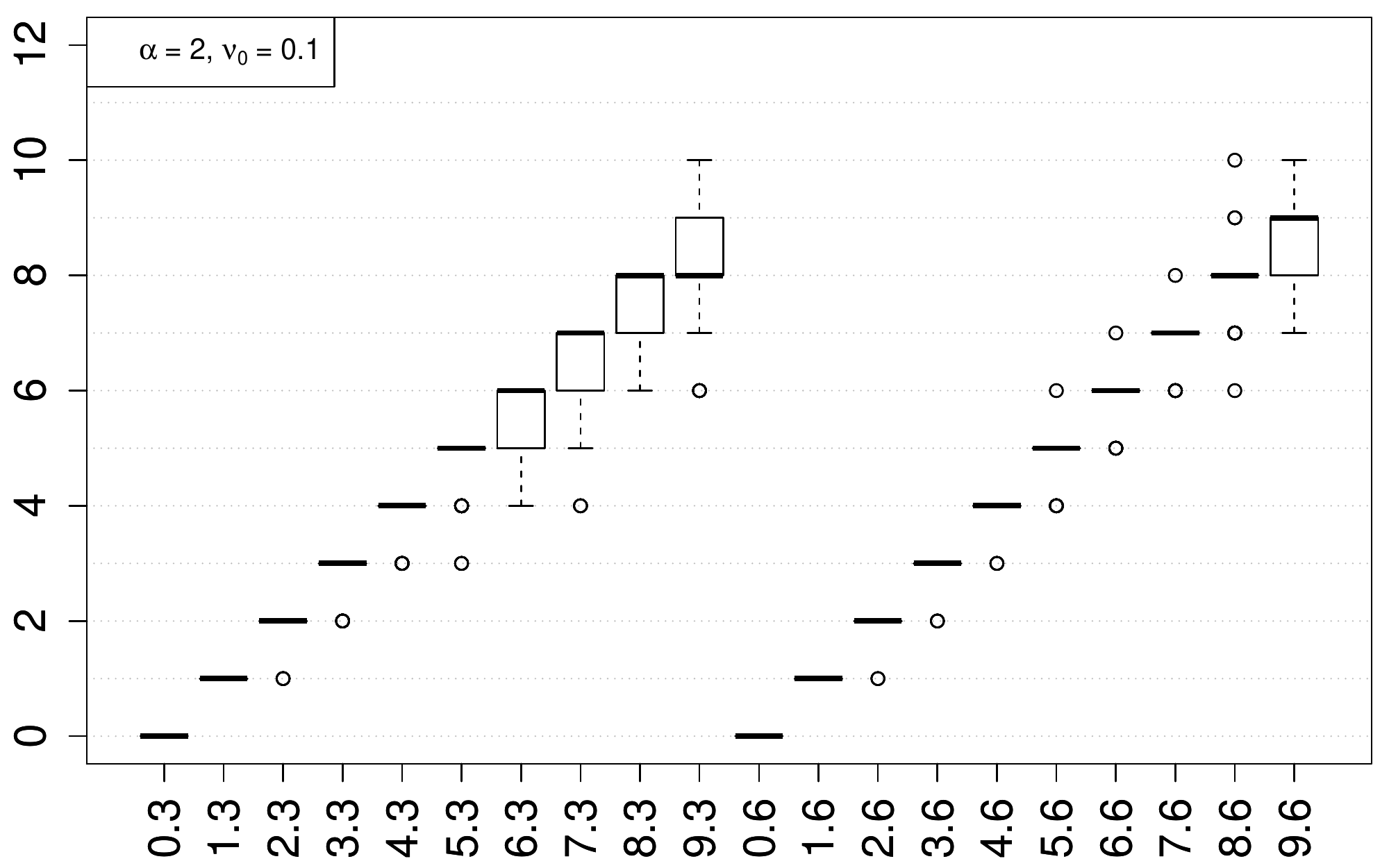}&
\includegraphics[scale=0.33]{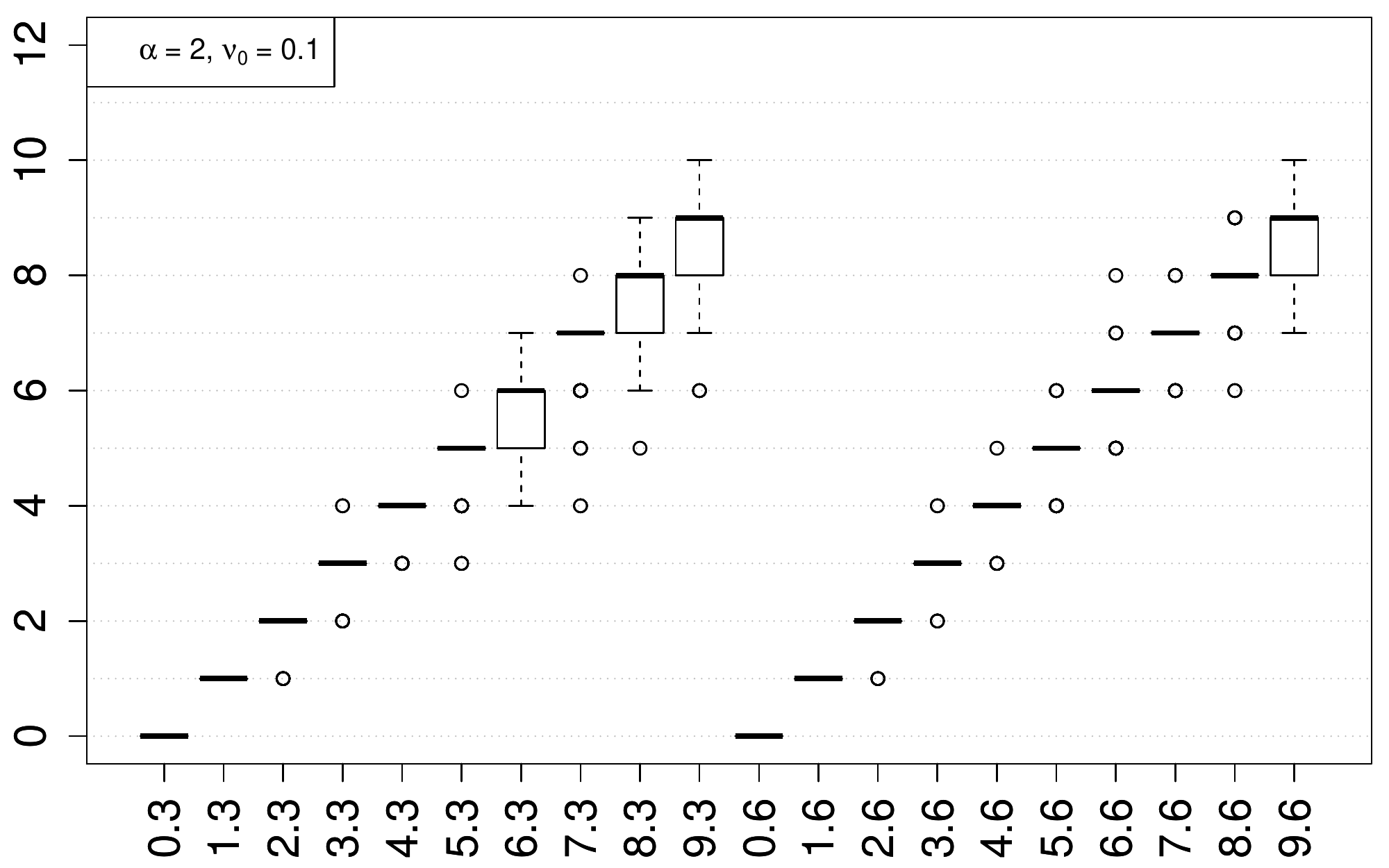}
\\
\includegraphics[scale=0.33]{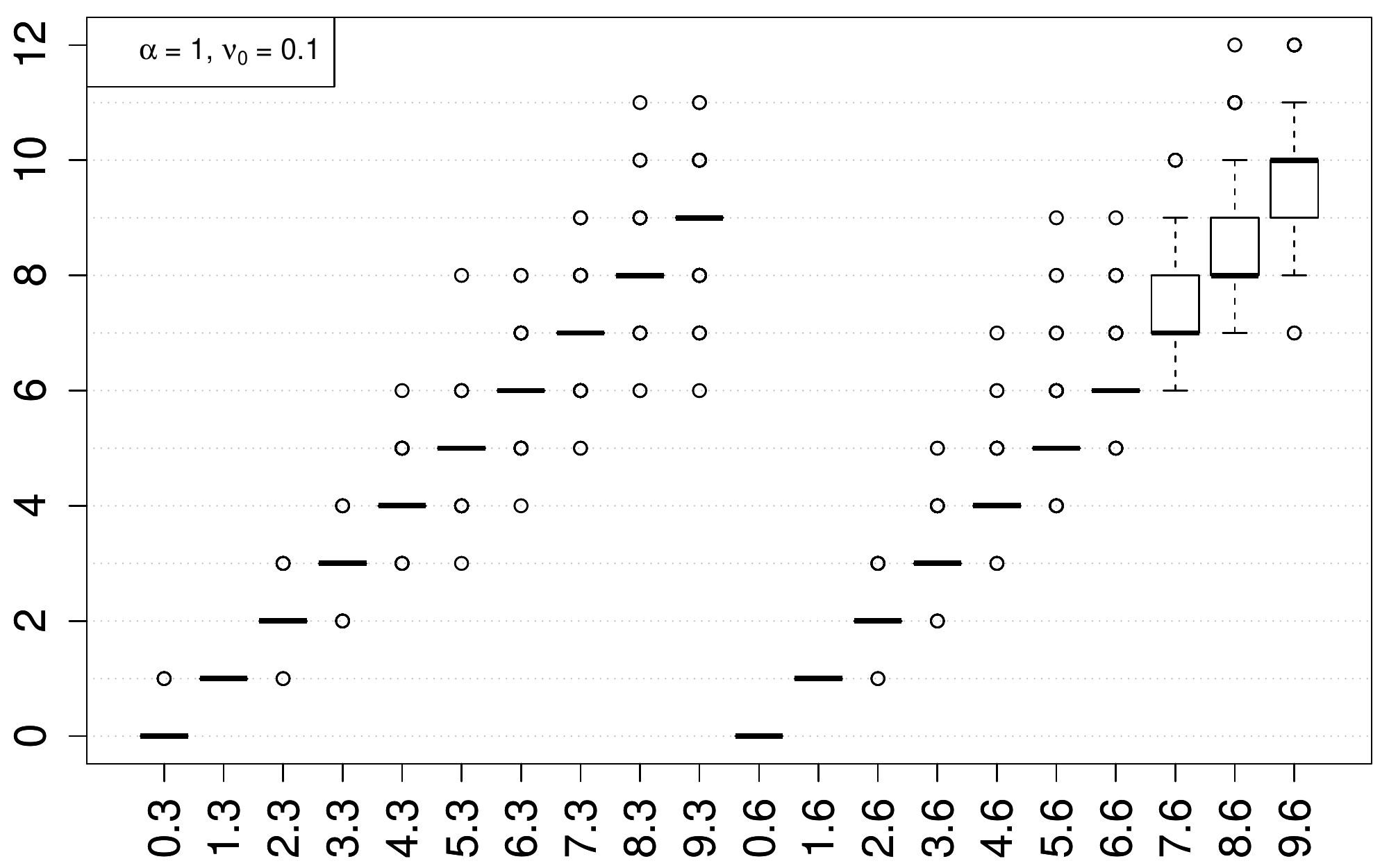}&
\includegraphics[scale=0.33]{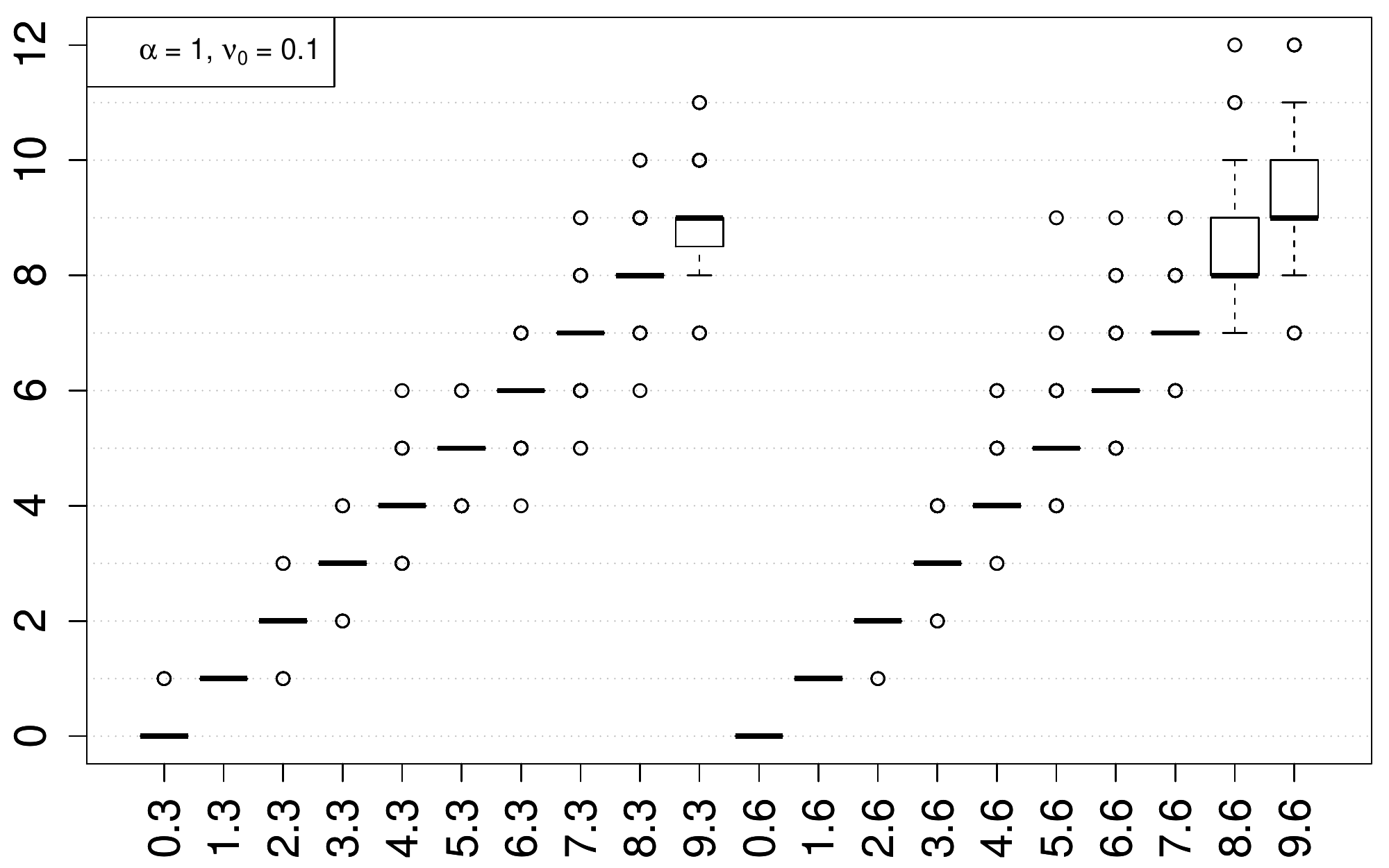}
\\
\includegraphics[scale=0.33]{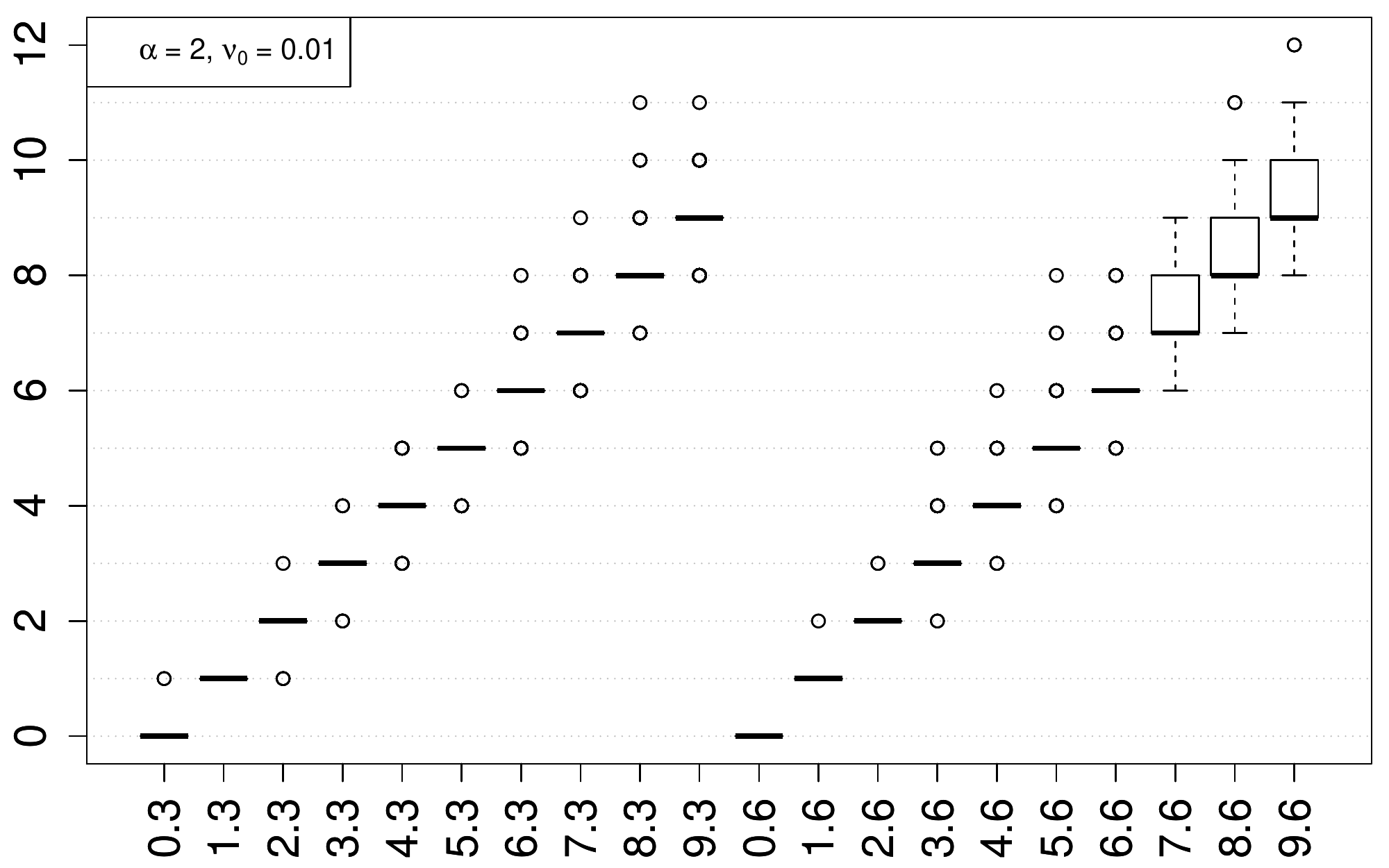}&
\includegraphics[scale=0.33]{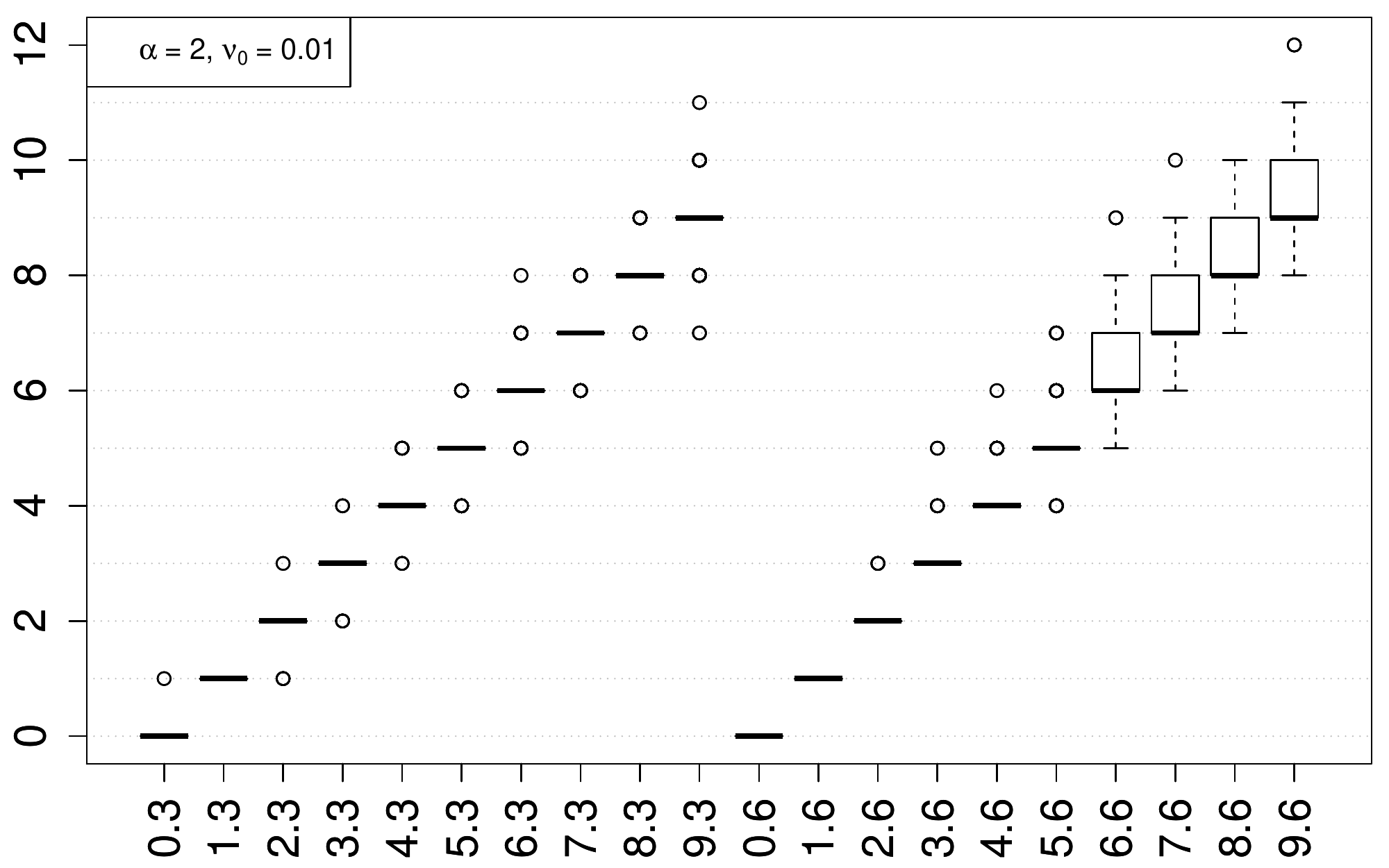}
\\
\includegraphics[scale=0.33]{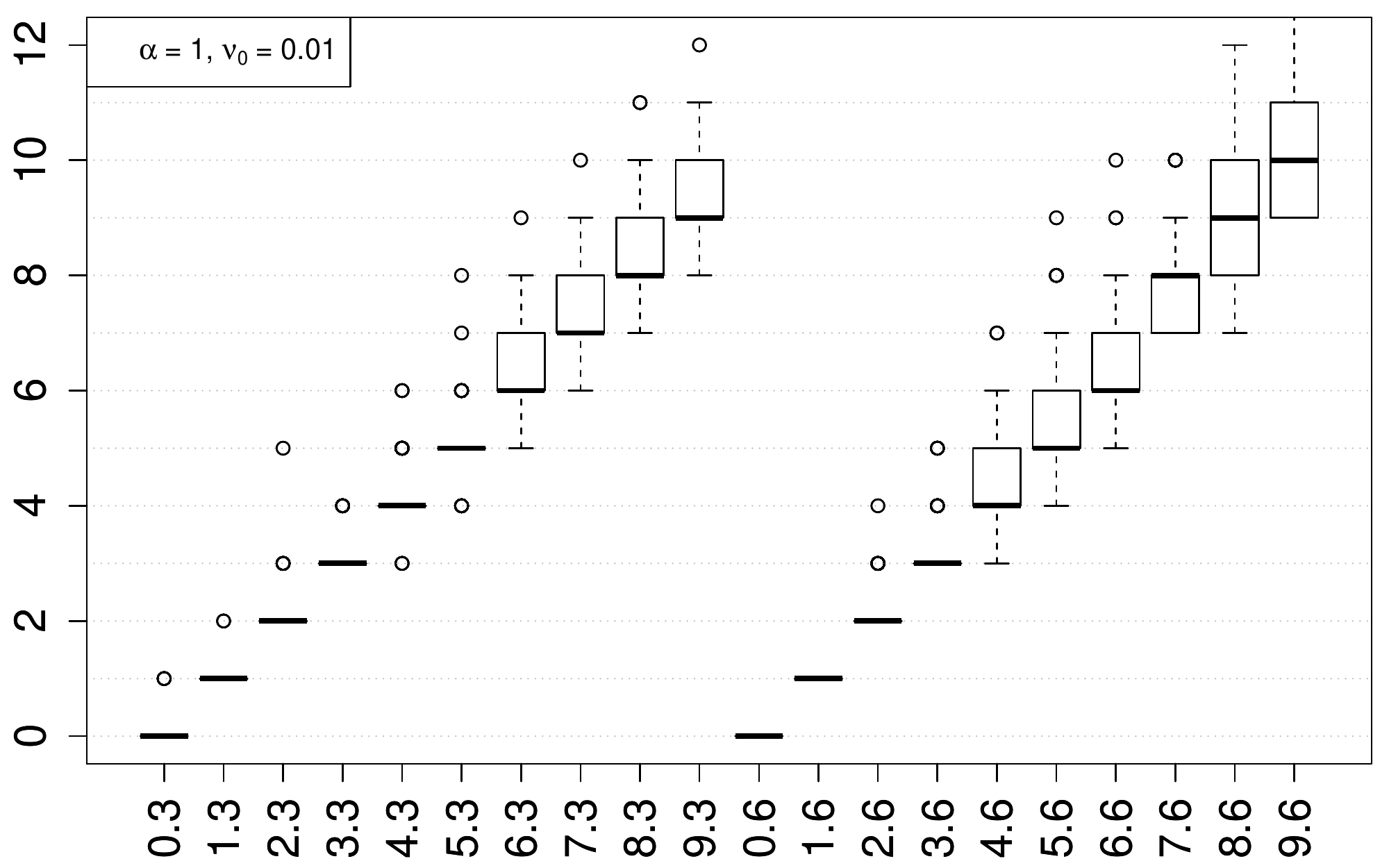}&
\includegraphics[scale=0.33]{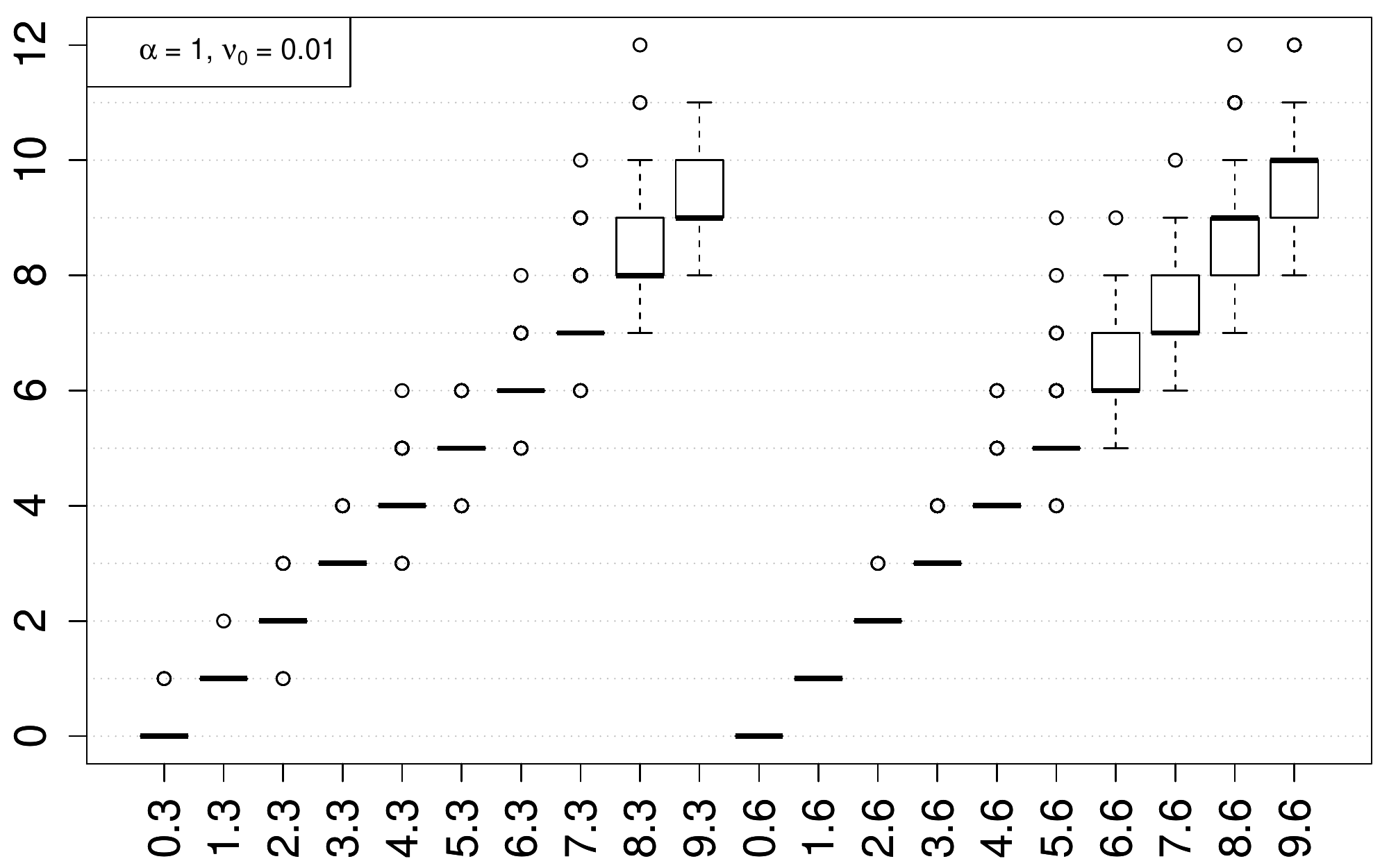}
\\
(a) Noisy Scenario & (b) Exact Scenario
\end{tabular}
\caption{Benchmarking the estimation of the number of change-points on synthetic datasets according to the ``noisy'' and ``exact'' simulation scenarios. Different combinations of prior parameters $\nu_0$ in Equation \eqref{eq:thetaPrior} and $\alpha$ in Equation \eqref{eq:ellPrior} were used to the MCMC sampler. Each pair of numbers in the horizontal axis displays the true number of change-points (first entry) and number of replicates (second entry).}
\label{fig:change-pointsEstimation}
\end{figure}

The estimate of the number of change-points are shown in Figure \ref{fig:change-pointsEstimation}.(a) and \ref{fig:change-pointsEstimation}.(b) for the noisy and exact simulation scenarios. We should note that the results are improved when replicating the analysis based on simulated datasets which are generated exactly by the assumed model, as illustrated in Figure \ref{fig:change-pointsEstimation}.(b), particularly when $\alpha = 1$ and $\nu_0 = 0.01$. 

\section{Benchmarking against the Poisson distribution on the real dataset}\label{Appsec:pois}

\begin{figure}
\begin{center}
\includegraphics[scale = 0.55]{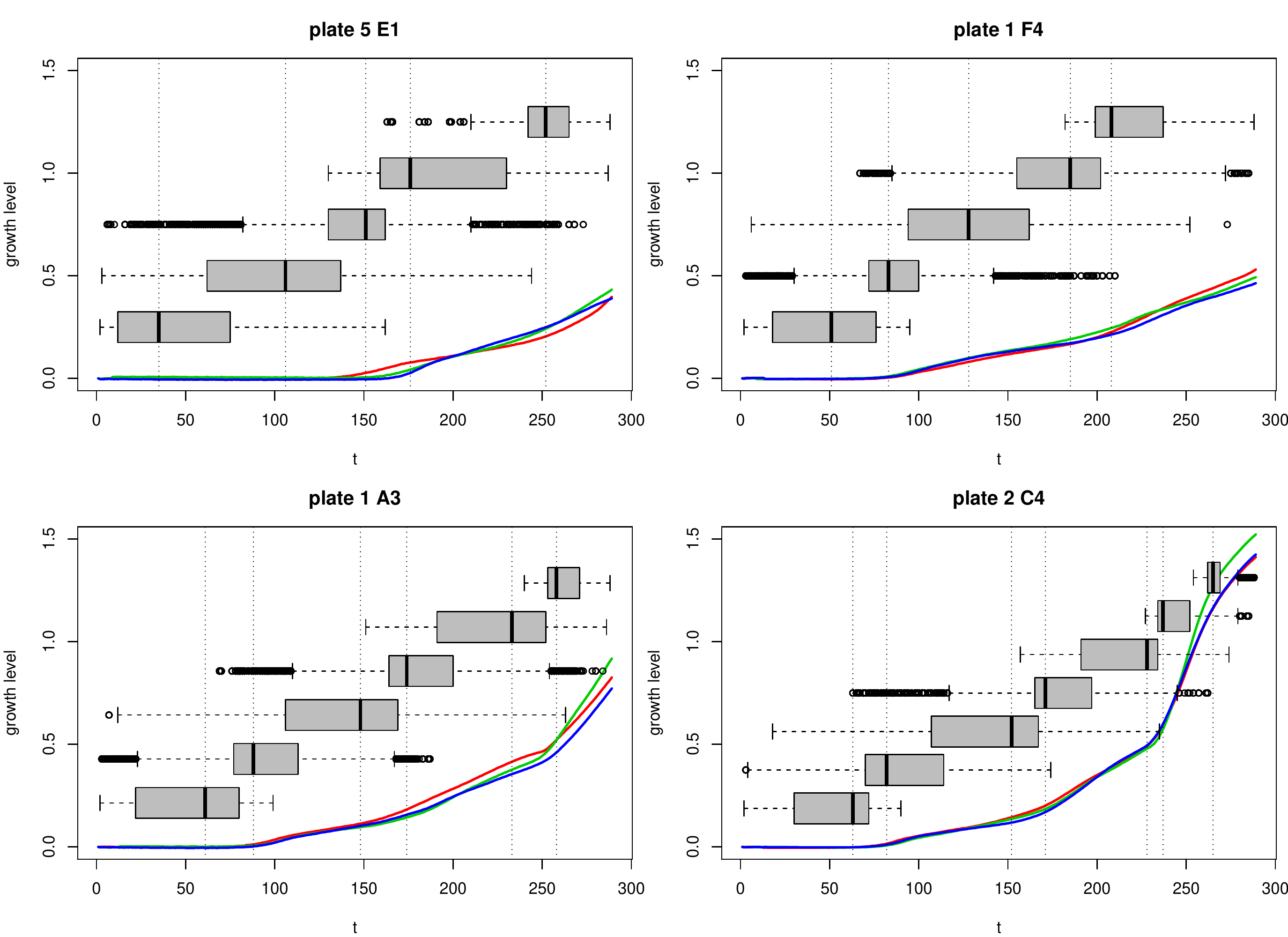}
\end{center}
\caption{Change-point locations conditionally on the MAP number of change-points according to a Poisson(1) prior distribution, truncated on the set $\{0,1,\ldots,30\}$ for four time-series of our real dataset.}
\label{fig:poisson}
\end{figure}

\begin{figure}
\begin{center}
\includegraphics[scale = 0.55]{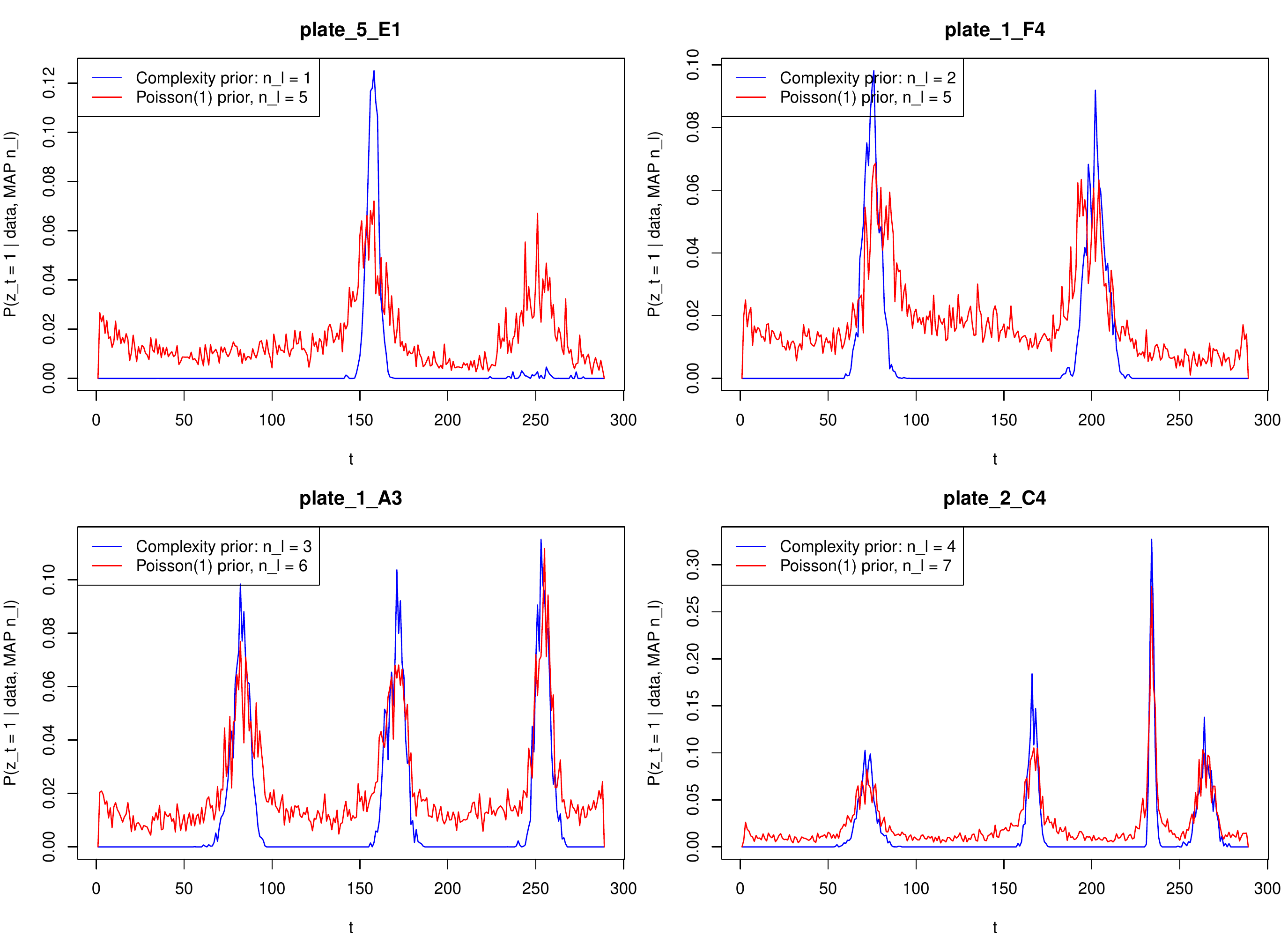}
\end{center}
\caption{Estimated posterior probabilities of the binary state variables $z_{nt}$, $t = 1,\ldots,T$ in Equation \ref{eq:z} for four time-series of the real dataset, conditionally on the most probable number of change-points ($n_\ell$), according to the complexity and Poisson(1) prior distributions on the number of change-points. The results are averaged across 20 independent chains.}
\label{fig:zProbs}
\end{figure}

Recall that in Section \ref{sec:sim} we demonstrated, using simulated data, that the estimated  number of change-points tends to overfit when using a Poisson instead the complexity prior distribution. Here, we illustrate that this is also the case on our real dataset. We consider again a Poisson(1) distribution, truncated on the set $\{0,1,\ldots,30\}$ and use the same four time-series of the real dataset, depicted in Figure 1 of the main paper. As shown in Figure \ref{fig:poisson}, the posterior distribution of the number of change-points under the Poisson prior distribution supports much larger values than the complexity prior. The estimated posterior mode of the number of change-points correspond to 5 for ``plate 5 E1'' (instead of 1 under the complexity prior), 5 for ``plate 1 F4'' (instead of 2), 6 for ``plate 1 A3'' (instead of 3) and 7 for ``plate 2 C4'' (instead of 4). We conclude that in this case the sampler assigns additional change-points to  intermediate observation periods, compared to the ones selected under the complexity prior. These additional change-points identify very small changes in the slope of the time-series and they do not contribute much in the interpretation of growth characteristics.

In order to further inspect the differences between the results arising from the two different prior assumptions, let us define the following random variables:
\begin{equation}\label{eq:z}
z_{nt} = \begin{cases}
1, & \mbox{if } t\in\{\tau_1,\ldots,\tau_{\ell_n}\}\\
0, & \mbox{otherwise}
\end{cases}
\end{equation}
for $t=1,\ldots,T$, $n=1,\ldots,N$. Note that $z_{nt}$ is a binary random variable with 1 denoting the event that a change-point is assigned at time-point $t$, $t = 1,\ldots,T$, for time-series $n = 1,\ldots,N$. The posterior probability $P(z_{nt} = 1|\boldsymbol{x}, \ell_n)$ can be estimated directly by averaging across the MCMC output. Figure \ref{fig:zProbs} illustrates the estimates of these posterior probabilities for a subset of four time-series. The peaks correspond to the locations of sampled change-points across the MCMC run. Notice that a blue peak is always accompanied by a red one, which means that under the Poisson distribution the set of inferred change-points actually contains the change-points locations selected from the complexity prior. However, the reverse is not necessarily true, especially for the first time series ("plate 5 E1"). Finally, notice that under the complexity prior distribution all intermediate time-points between two peaks are assigned  zero posterior probabilities of containing change-points. This is not the case for the Poisson prior distribution, where all time-points contain a change-point with strictly positive posterior probabilities, which is due to the presence of the additional change-points.

\section{Additional benchmarking against  the narrowest-over-threshold method}\label{Appsec:not}

\begin{figure}
\begin{center}
\includegraphics[scale = 0.55]{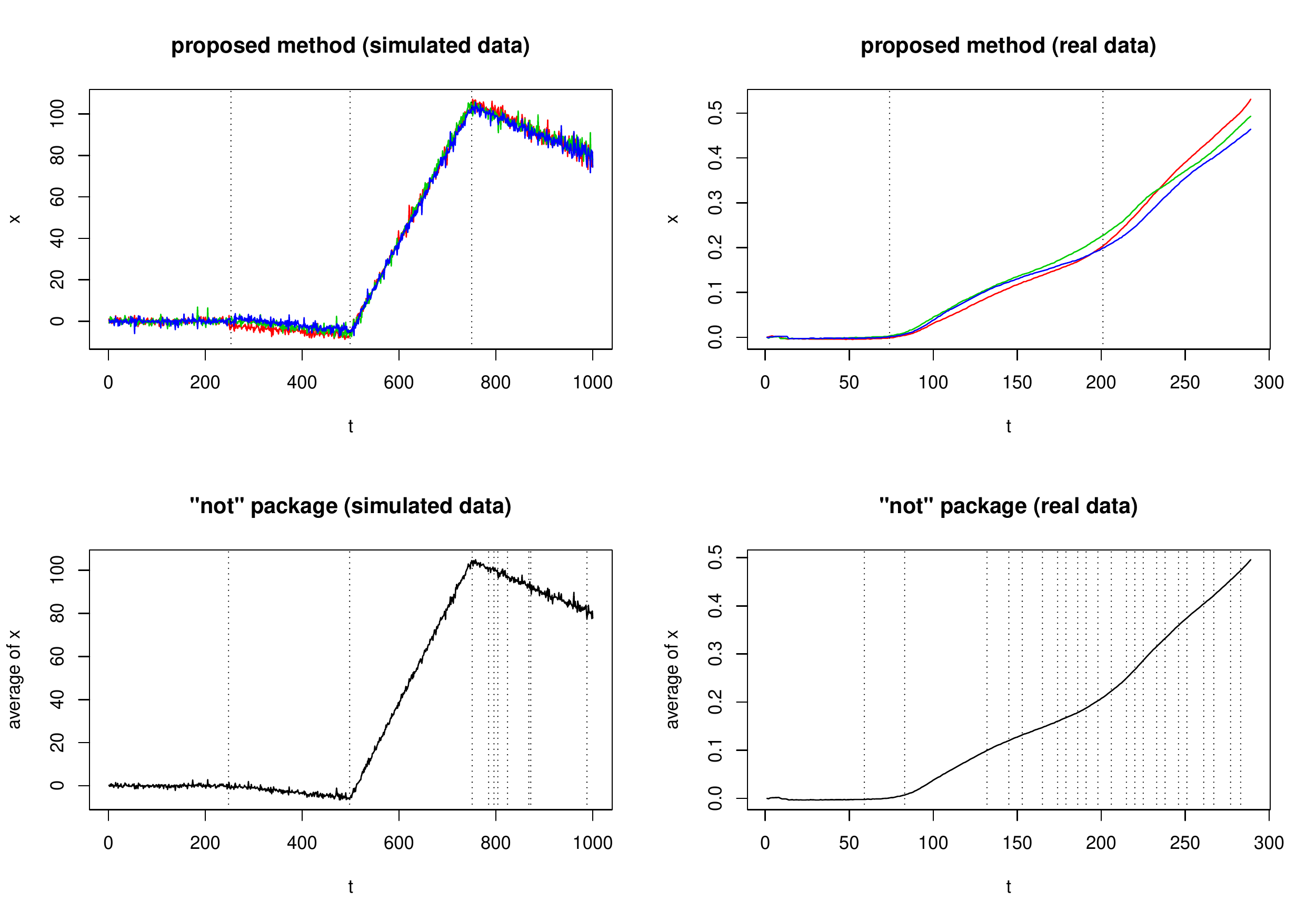}
\end{center}
\caption{Comparison with the narrowest-over-threshold method as implemented in the {\tt not} R package. Inferred change-points are indicated by dotted lines.}
\label{fig:not}
\end{figure}

In this section we provide some additional results regarding synthetic and true datasets when using the {\tt not} package of \cite{baranowski2016narrowest}. We consider one of our simulated datasets as well as one time-series of our real data, which illustrate the typical performance of each method in our data.  As already mentioned, {\tt not} deals with univariate time-series so in order to apply this method we averaged across the replicates of the time-series. The results are illustrated in Figure \ref{fig:not}. The first column consists of a simulated dataset (using the same generating mechanism described in Section \ref{Appsec:sim}, where the number of change-points is equal to 3 and our proposed method  is able to succesfully detect it (shown in first row of Figure \ref{fig:not}). On the other hand,  the method of \cite{baranowski2016narrowest} overfits the number of change-points. A similar pattern is illustrated for the real dataset shown in the second column of \ref{fig:not}, where the narrowest-over-threshold method infers a much larger number of change-points compared to our approach, which is not realistic.

\bibliography{main}

\begin{thebibliography}{46}
\providecommand{\natexlab}[1]{#1}
\providecommand{\url}[1]{\texttt{#1}}
\expandafter\ifx\csname urlstyle\endcsname\relax
  \providecommand{\doi}[1]{doi: #1}\else
  \providecommand{\doi}{doi: \begingroup \urlstyle{rm}\Url}\fi

\bibitem[Amich et~al.(2013)Amich, Schafferer, Haas, and
  Krappmann]{10.1371/journal.ppat.1003573}
J.~Amich, L.~Schafferer, H.~Haas, and S.~Krappmann.
\newblock Regulation of sulphur assimilation is essential for virulence and
  affects iron homeostasis of the human-pathogenic mould {A}spergillus
  fumigatus.
\newblock \emph{PLOS Pathogens}, 9\penalty0 (8):\penalty0 1--24, 08 2013.
\newblock \doi{10.1371/journal.ppat.1003573}.
\newblock URL \url{https://doi.org/10.1371/journal.ppat.1003573}.

\bibitem[Baranowski et~al.(2016)Baranowski, Chen, and
  Fryzlewicz]{baranowski2016narrowest}
R.~Baranowski, Y.~Chen, and P.~Fryzlewicz.
\newblock Narrowest-over-threshold detection of multiple change-points and
  change-point-like features.
\newblock \emph{arXiv preprint arXiv:1609.00293}, 2016.

\bibitem[Basseville et~al.(1993)Basseville, Nikiforov,
  et~al.]{basseville1993detection}
M.~Basseville, I.~V. Nikiforov, et~al.
\newblock \emph{Detection of abrupt changes: theory and application}, volume
  104.
\newblock Prentice Hall Englewood Cliffs, 1993.

\bibitem[Bertuzzi et~al.(2014)Bertuzzi, Schrettl, Alcazar-Fuoli, Cairns,
  Mu{\~n}oz, Walker, Herbst, Safari, Cheverton, Chen, et~al.]{bertuzzi2014ph}
M.~Bertuzzi, M.~Schrettl, L.~Alcazar-Fuoli, T.~C. Cairns, A.~Mu{\~n}oz, L.~A.
  Walker, S.~Herbst, M.~Safari, A.~M. Cheverton, D.~Chen, et~al.
\newblock The p{H}-responsive {P}ac{C} transcription factor of {A}spergillus
  fumigatus governs epithelial entry and tissue invasion during pulmonary
  aspergillosis.
\newblock \emph{PLoS pathogens}, 10\penalty0 (10):\penalty0 e1004413, 2014.

\bibitem[Brown et~al.(2012)Brown, Denning, Gow, Levitz, Netea, and
  White]{brown2012hidden}
G.~D. Brown, D.~W. Denning, N.~A. Gow, S.~M. Levitz, M.~G. Netea, and T.~C.
  White.
\newblock Hidden killers: human fungal infections.
\newblock \emph{Science translational medicine}, 4\penalty0 (165):\penalty0
  165rv13--165rv13, 2012.

\bibitem[Cahill et~al.(2015)Cahill, Rahmstorf, and
  Parnell]{1748-9326-10-8-084002}
N.~Cahill, S.~Rahmstorf, and A.~C. Parnell.
\newblock Change points of global temperature.
\newblock \emph{Environmental Research Letters}, 10\penalty0 (8):\penalty0
  084002, 2015.
\newblock URL \url{http://stacks.iop.org/1748-9326/10/i=8/a=084002}.

\bibitem[Castillo and van~der Vaart(2012)]{castillo2012}
I.~Castillo and A.~van~der Vaart.
\newblock Needles and straw in a haystack: {P}osterior concentration for
  possibly sparse sequences.
\newblock \emph{The Annals of Statistics}, 40\penalty0 (4):\penalty0
  2069--2101, 08 2012.
\newblock \doi{10.1214/12-AOS1029}.
\newblock URL \url{http://dx.doi.org/10.1214/12-AOS1029}.

\bibitem[Chamroukhi et~al.(2013)Chamroukhi, Mohammed, Trabelsi, Oukhellou, and
  Amirat]{Chamroukhi2013633}
F.~Chamroukhi, S.~Mohammed, D.~Trabelsi, L.~Oukhellou, and Y.~Amirat.
\newblock Joint segmentation of multivariate time series with hidden process
  regression for human activity recognition.
\newblock \emph{Neurocomputing}, 120:\penalty0 633 -- 644, 2013.
\newblock ISSN 0925-2312.
\newblock \doi{https://doi.org/10.1016/j.neucom.2013.04.003}.
\newblock URL
  \url{http://www.sciencedirect.com/science/article/pii/S0925231213004086}.
\newblock Image Feature Detection and Description.

\bibitem[Chib(1995)]{chib1995marginal}
S.~Chib.
\newblock Marginal likelihood from the {G}ibbs output.
\newblock \emph{Journal of the American Statistical Association}, 90\penalty0
  (432):\penalty0 1313--1321, 1995.

\bibitem[Chib(1998)]{Chib1998221}
S.~Chib.
\newblock Estimation and comparison of multiple change-point models.
\newblock \emph{Journal of Econometrics}, 86\penalty0 (2):\penalty0 221 -- 241,
  1998.
\newblock ISSN 0304-4076.
\newblock \doi{https://doi.org/10.1016/S0304-4076(97)00115-2}.
\newblock URL
  \url{http://www.sciencedirect.com/science/article/pii/S0304407697001152}.

\bibitem[Dinamarco et~al.(2012)Dinamarco, Almeida, de~Castro, Brown, dos Reis,
  Ramalho, Savoldi, Goldman, and Goldman]{dinamarco2012molecular}
T.~M. Dinamarco, R.~S. Almeida, P.~A. de~Castro, N.~A. Brown, T.~F. dos Reis,
  L.~N.~Z. Ramalho, M.~Savoldi, M.~H.~S. Goldman, and G.~H. Goldman.
\newblock Molecular characterization of the putative transcription factor
  {S}eb{A} involved in virulence in {A}spergillus fumigatus.
\newblock \emph{Eukaryotic cell}, 11\penalty0 (4):\penalty0 518--531, 2012.

\bibitem[Dobigeon et~al.(2007)Dobigeon, Tourneret, and
  Scargle]{dobigeon2007joint}
N.~Dobigeon, J.-Y. Tourneret, and J.~D. Scargle.
\newblock Joint segmentation of multivariate astronomical time series:
  {B}ayesian sampling with a hierarchical model.
\newblock \emph{IEEE Transactions on Signal Processing}, 55\penalty0
  (2):\penalty0 414--423, 2007.

\bibitem[Fearnhead(2006)]{fearnhead2006exact}
P.~Fearnhead.
\newblock Exact and efficient {B}ayesian inference for multiple changepoint
  problems.
\newblock \emph{Statistics and computing}, 16\penalty0 (2):\penalty0 203--213,
  2006.

\bibitem[Fearnhead et~al.(2018)Fearnhead, Maidstone, and
  Letchford]{maidstone2017detecting}
P.~Fearnhead, R.~Maidstone, and A.~Letchford.
\newblock Detecting changes in slope with an l0 penalty.
\newblock \emph{Journal of Computational and Graphical Statistics}, 0\penalty0
  (0):\penalty0 1--11, 2018.
\newblock \doi{10.1080/10618600.2018.1512868}.
\newblock URL \url{https://doi.org/10.1080/10618600.2018.1512868}.

\bibitem[Fischer and Sawers(2013)]{fischer2013universally}
M.~Fischer and R.~G. Sawers.
\newblock A universally applicable and rapid method for measuring the growth of
  {S}treptomyces and other filamentous microorganisms by methylene blue
  adsorption-desorption.
\newblock \emph{Applied and environmental microbiology}, 79\penalty0
  (14):\penalty0 4499--4502, 2013.

\bibitem[Frick et~al.(2014)Frick, Munk, and Sieling]{RSSB:RSSB12047}
K.~Frick, A.~Munk, and H.~Sieling.
\newblock Multiscale change point inference.
\newblock \emph{Journal of the Royal Statistical Society: Series B (Statistical
  Methodology)}, 76\penalty0 (3):\penalty0 495--580, 2014.
\newblock ISSN 1467-9868.
\newblock \doi{10.1111/rssb.12047}.
\newblock URL \url{http://dx.doi.org/10.1111/rssb.12047}.

\bibitem[Fryzlewicz et~al.(2014)]{fryzlewicz2014wild}
P.~Fryzlewicz et~al.
\newblock Wild binary segmentation for multiple change-point detection.
\newblock \emph{The Annals of Statistics}, 42\penalty0 (6):\penalty0
  2243--2281, 2014.

\bibitem[Green(1995)]{green1995reversible}
P.~J. Green.
\newblock Reversible jump {M}arkov chain {M}onte {C}arlo computation and
  {B}ayesian model determination.
\newblock \emph{Biometrika}, 82\penalty0 (4):\penalty0 711--732, 1995.

\bibitem[Gsaller et~al.(2016)Gsaller, Hortschansky, Furukawa, Carr, Rash,
  Capilla, Müller, Bracher, Bowyer, Haas, Brakhage, and
  Bromley]{10.1371/journal.ppat.1005775}
F.~Gsaller, P.~Hortschansky, T.~Furukawa, P.~D. Carr, B.~Rash, J.~Capilla,
  C.~Müller, F.~Bracher, P.~Bowyer, H.~Haas, A.~A. Brakhage, and M.~J.
  Bromley.
\newblock Sterol biosynthesis and azole tolerance is governed by the opposing
  actions of {S}rb{A} and the {C}{C}{A}{A}{T} binding complex.
\newblock \emph{PLOS Pathogens}, 12\penalty0 (7):\penalty0 1--22, 07 2016.
\newblock \doi{10.1371/journal.ppat.1005775}.
\newblock URL \url{https://doi.org/10.1371/journal.ppat.1005775}.

\bibitem[Halpern(2000)]{BIOM:BIOM903}
A.~L. Halpern.
\newblock Multiple-changepoint testing for an alternating segments model of a
  binary sequence.
\newblock \emph{Biometrics}, 56\penalty0 (3):\penalty0 903--908, 2000.
\newblock ISSN 1541-0420.
\newblock \doi{10.1111/j.0006-341X.2000.00903.x}.
\newblock URL \url{http://dx.doi.org/10.1111/j.0006-341X.2000.00903.x}.

\bibitem[Hastings(1970)]{10.2307/2334940}
W.~K. Hastings.
\newblock Monte {C}arlo sampling methods using {M}arkov chains and their
  applications.
\newblock \emph{Biometrika}, 57\penalty0 (1):\penalty0 97--109, 1970.
\newblock ISSN 00063444.
\newblock URL \url{http://www.jstor.org/stable/2334940}.

\bibitem[He(2017)]{doi:10.1080/03610926.2016.1161797}
C.~He.
\newblock Bayesian multiple change-point estimation for exponential
  distribution with truncated and censored data.
\newblock \emph{Communications in Statistics - Theory and Methods}, 46\penalty0
  (12):\penalty0 5827--5839, 2017.
\newblock \doi{10.1080/03610926.2016.1161797}.
\newblock URL \url{http://dx.doi.org/10.1080/03610926.2016.1161797}.

\bibitem[Hutter(2007)]{hutter2007}
M.~Hutter.
\newblock Exact {B}ayesian regression of piecewise constant functions.
\newblock \emph{Bayesian Anal.}, 2\penalty0 (4):\penalty0 635--664, 12 2007.
\newblock \doi{10.1214/07-BA225}.
\newblock URL \url{http://dx.doi.org/10.1214/07-BA225}.

\bibitem[Jackson et~al.(2005)Jackson, Scargle, Barnes, Arabhi, Alt, Gioumousis,
  Gwin, Sangtrakulcharoen, Tan, and Tsai]{jackson2005algorithm}
B.~Jackson, J.~D. Scargle, D.~Barnes, S.~Arabhi, A.~Alt, P.~Gioumousis,
  E.~Gwin, P.~Sangtrakulcharoen, L.~Tan, and T.~T. Tsai.
\newblock An algorithm for optimal partitioning of data on an interval.
\newblock \emph{IEEE Signal Processing Letters}, 12\penalty0 (2):\penalty0
  105--108, 2005.

\bibitem[Johnson et~al.(2003)Johnson, Elashoff, and
  Harkema]{doi:10.1093/biostatistics/4.1.143}
T.~D. Johnson, R.~M. Elashoff, and S.~J. Harkema.
\newblock A {B}ayesian change‐point analysis of electromyographic data:
  detecting muscle activation patterns and associated applications.
\newblock \emph{Biostatistics}, 4\penalty0 (1):\penalty0 143, 2003.
\newblock \doi{10.1093/biostatistics/4.1.143}.
\newblock URL \url{+ http://dx.doi.org/10.1093/biostatistics/4.1.143}.

\bibitem[Killick et~al.(2012)Killick, Fearnhead, and
  Eckley]{killick2012optimal}
R.~Killick, P.~Fearnhead, and I.~A. Eckley.
\newblock Optimal detection of changepoints with a linear computational cost.
\newblock \emph{Journal of the American Statistical Association}, 107\penalty0
  (500):\penalty0 1590--1598, 2012.

\bibitem[Kim and Cheon(2010)]{Kim2010}
J.~Kim and S.~Cheon.
\newblock Bayesian multiple change-point estimation with annealing stochastic
  approximation {M}onte {C}arlo.
\newblock \emph{Computational Statistics}, 25\penalty0 (2):\penalty0 215--239,
  2010.
\newblock ISSN 1613-9658.
\newblock \doi{10.1007/s00180-009-0172-x}.
\newblock URL \url{http://dx.doi.org/10.1007/s00180-009-0172-x}.

\bibitem[Lavielle and Lebarbier(2001)]{lavielle2001application}
M.~Lavielle and E.~Lebarbier.
\newblock An application of {M}{C}{M}{C} methods for the multiple change-points
  problem.
\newblock \emph{Signal Processing}, 81\penalty0 (1):\penalty0 39--53, 2001.

\bibitem[Lee et~al.(2016)Lee, Kwon, Lee, Jung, Kim, and Yu]{lee2016negative}
M.-K. Lee, N.-J. Kwon, I.-S. Lee, S.~Jung, S.-C. Kim, and J.-H. Yu.
\newblock Negative regulation and developmental competence in {A}spergillus.
\newblock \emph{Scientific reports}, 6:\penalty0 28874, 2016.
\newblock \doi{10.1038/srep28874}.

\bibitem[Liu et~al.(1997)Liu, Wu, and Zidek]{liu1997segmented}
J.~Liu, S.~Wu, and J.~V. Zidek.
\newblock On segmented multivariate regression.
\newblock \emph{Statistica Sinica}, pages 497--525, 1997.

\bibitem[Lu et~al.(2010)Lu, Lund, and Lee]{lu2010}
Q.~Lu, R.~Lund, and T.~C.~M. Lee.
\newblock An {M}{D}{L} approach to the climate segmentation problem.
\newblock \emph{Ann. Appl. Stat.}, 4\penalty0 (1):\penalty0 299--319, 03 2010.
\newblock \doi{10.1214/09-AOAS289}.
\newblock URL \url{http://dx.doi.org/10.1214/09-AOAS289}.

\bibitem[Metropolis et~al.(1953)Metropolis, Rosenbluth, Rosenbluth, Teller, and
  Teller]{metropolis1953equation}
N.~Metropolis, A.~W. Rosenbluth, M.~N. Rosenbluth, A.~H. Teller, and E.~Teller.
\newblock Equation of state calculations by fast computing machines.
\newblock \emph{The journal of chemical physics}, 21\penalty0 (6):\penalty0
  1087--1092, 1953.

\bibitem[Papastamoulis(2017)]{beast}
P.~Papastamoulis.
\newblock \emph{beast: Bayesian Estimation of Change-Points in the Slope of
  Multivariate Time-Series}, 2017.
\newblock URL \url{http://CRAN.R-project.org/package=beast}.
\newblock R package version 1.0.

\bibitem[Picard et~al.(2011)Picard, Lebarbier, Budinskà, and
  Robin]{Picard20111160}
F.~Picard, E.~Lebarbier, E.~Budinskà, and S.~Robin.
\newblock Joint segmentation of multivariate {G}aussian processes using mixed
  linear models.
\newblock \emph{Computational Statistics \& Data Analysis}, 55\penalty0
  (2):\penalty0 1160 -- 1170, 2011.
\newblock ISSN 0167-9473.
\newblock \doi{https://doi.org/10.1016/j.csda.2010.09.015}.
\newblock URL
  \url{http://www.sciencedirect.com/science/article/pii/S0167947310003580}.

\bibitem[Punskaya et~al.(2002)Punskaya, Andrieu, Doucet, and
  Fitzgerald]{984776}
E.~Punskaya, C.~Andrieu, A.~Doucet, and W.~J. Fitzgerald.
\newblock Bayesian curve fitting using {M}{C}{M}{C} with applications to signal
  segmentation.
\newblock \emph{IEEE Transactions on Signal Processing}, 50\penalty0
  (3):\penalty0 747--758, Mar 2002.
\newblock ISSN 1053-587X.
\newblock \doi{10.1109/78.984776}.

\bibitem[{R Development Core Team}(2008)]{rcitation}
{R Development Core Team}.
\newblock \emph{R: A Language and Environment for Statistical Computing}.
\newblock R Foundation for Statistical Computing, Vienna, Austria, 2008.
\newblock URL \url{http://www.R-project.org}.
\newblock {ISBN} 3-900051-07-0.

\bibitem[Rudoy et~al.(2010)Rudoy, Yuen, Howe, and Wolfe]{RSSC:RSSC715}
D.~Rudoy, S.~G. Yuen, R.~D. Howe, and P.~J. Wolfe.
\newblock Bayesian change-point analysis for atomic force microscopy and soft
  material indentation.
\newblock \emph{Journal of the Royal Statistical Society: Series C (Applied
  Statistics)}, 59\penalty0 (4):\penalty0 573--593, 2010.
\newblock ISSN 1467-9876.
\newblock \doi{10.1111/j.1467-9876.2010.00715.x}.
\newblock URL \url{http://dx.doi.org/10.1111/j.1467-9876.2010.00715.x}.

\bibitem[Schroeder and Fryzlewicz(2013)]{Schroeder}
A.~L. Schroeder and P.~Fryzlewicz.
\newblock Adaptive trend estimation in financial time series via multiscale
  change-point-induced basis recovery.
\newblock \emph{Statistics and its interface}, 6\penalty0 (4):\penalty0
  449--461, 2013.

\bibitem[Sch\"utz and Holschneider(2011)]{PhysRevE.84.021120}
N.~Sch\"utz and M.~Holschneider.
\newblock Detection of trend changes in time series using {B}ayesian inference.
\newblock \emph{Phys. Rev. E}, 84:\penalty0 021120, Aug 2011.
\newblock \doi{10.1103/PhysRevE.84.021120}.
\newblock URL \url{https://link.aps.org/doi/10.1103/PhysRevE.84.021120}.

\bibitem[Schwaller and Robin(2017)]{Schwaller2017}
L.~Schwaller and S.~Robin.
\newblock Exact bayesian inference for off-line change-point detection in
  tree-structured graphical models.
\newblock \emph{Statistics and Computing}, 27\penalty0 (5):\penalty0
  1331--1345, Sep 2017.
\newblock ISSN 1573-1375.
\newblock \doi{10.1007/s11222-016-9689-3}.
\newblock URL \url{https://doi.org/10.1007/s11222-016-9689-3}.

\bibitem[Scott and Knott(1974)]{10.2307/2529204}
A.~J. Scott and M.~Knott.
\newblock A cluster analysis method for grouping means in the analysis of
  variance.
\newblock \emph{Biometrics}, 30\penalty0 (3):\penalty0 507--512, 1974.
\newblock ISSN 0006341X, 15410420.
\newblock URL \url{http://www.jstor.org/stable/2529204}.

\bibitem[Stevenson et~al.(2016)Stevenson, McVey, Clark, Swain, and
  Pilizota]{stevenson2016general}
K.~Stevenson, A.~F. McVey, I.~B. Clark, P.~S. Swain, and T.~Pilizota.
\newblock General calibration of microbial growth in microplate readers.
\newblock \emph{Scientific reports}, 6:\penalty0 38828, 2016.

\bibitem[Tai et~al.(2010)Tai, Kvale, and Witte]{BIOM:BIOM1328}
Y.~C. Tai, M.~N. Kvale, and J.~S. Witte.
\newblock Segmentation and estimation for {S}{N}{P} microarrays: A {B}ayesian
  multiple change-point approach.
\newblock \emph{Biometrics}, 66\penalty0 (3):\penalty0 675--683, 2010.
\newblock ISSN 1541-0420.
\newblock \doi{10.1111/j.1541-0420.2009.01328.x}.
\newblock URL \url{http://dx.doi.org/10.1111/j.1541-0420.2009.01328.x}.

\bibitem[Willger et~al.(2008)Willger, Puttikamonkul, Kim, Burritt, Grahl,
  Metzler, Barbuch, Bard, Lawrence, and Cramer]{10.1371}
S.~D. Willger, S.~Puttikamonkul, K.-H. Kim, J.~B. Burritt, N.~Grahl, L.~J.
  Metzler, R.~Barbuch, M.~Bard, C.~B. Lawrence, and R.~A. Cramer, Jr.
\newblock A sterol-regulatory element binding protein is required for cell
  polarity, hypoxia adaptation, azole drug resistance, and virulence in
  {A}spergillus fumigatus.
\newblock \emph{PLOS Pathogens}, 4\penalty0 (11):\penalty0 1--18, 11 2008.
\newblock \doi{10.1371/journal.ppat.1000200}.
\newblock URL \url{https://doi.org/10.1371/journal.ppat.1000200}.

\bibitem[Yildirim et~al.(2013)Yildirim, Singh, and
  Doucet]{doi:10.1080/10618600.2012.674653}
S.~Yildirim, S.~S. Singh, and A.~Doucet.
\newblock An online expectation--maximization algorithm for changepoint models.
\newblock \emph{Journal of Computational and Graphical Statistics}, 22\penalty0
  (4):\penalty0 906--926, 2013.
\newblock \doi{10.1080/10618600.2012.674653}.
\newblock URL \url{http://dx.doi.org/10.1080/10618600.2012.674653}.

\bibitem[Zhao and Chu(2010)]{zhao2010bayesian}
X.~Zhao and P.-S. Chu.
\newblock Bayesian changepoint analysis for extreme events (typhoons, heavy
  rainfall, and heat waves): An {R}{J}{M}{C}{M}{C} approach.
\newblock \emph{Journal of Climate}, 23\penalty0 (5):\penalty0 1034--1046,
  2010.

\end{thebibliography}

\end{document}